\documentclass[11pt,twoside,american,english]{article}
\usepackage[T1]{fontenc}
\usepackage{geometry}
\usepackage{color}
\usepackage{babel}
\usepackage{float}
\usepackage{wrapfig}
\usepackage{amsmath}
\usepackage{amssymb}
\usepackage{graphicx}
\usepackage[unicode=true,pdfusetitle,
 bookmarks=true,bookmarksnumbered=false,bookmarksopen=false,
 breaklinks=false,pdfborder={0 0 0},pdfborderstyle={},backref=false,colorlinks=true]
 {hyperref}

\makeatletter
\usepackage{sidecap} 

\makeatother

\begin{document}
\global\long\def\thMy{\mathrm{th}}
\global\long\def\ext{\mathrm{ext}}
\global\long\def\intMy{\mathrm{int}}
\global\long\def\x{\boldsymbol{x}}
\global\long\def\w{\boldsymbol{w}}
\global\long\def\tr{\mathrm{T}}
\global\long\def\CV{\mathrm{CV}}
\global\long\def\wnorm{\left|\boldsymbol{w}\right|}

\title{\textbf{Balanced Excitation and Inhibition are Required for High-Capacity,
Noise-Robust Neuronal Selectivity} }

\date{{}}

\author{\textbf{Ran Rubin$^{\star}$}\\
Department of Neuroscience\\
Columbia University\\
 New York NY 10027 USA\\
\textbf{}\\
\textbf{L. F. Abbott}\\
  Department of Neuroscience\\
 Department of Physiology and Cellular Biophysics\\
 Columbia University\\
 New York NY 10027 USA\\
\\
 and\\
\\
 \textbf{Haim Sompolinsky}\\
 Edmond and Lily Safra Center for Brain Sciences\\
 Racah Institute of Physics\\
 The Hebrew University of Jerusalem,\\
Edmond J. Safra Campus\\
 Jerusalem 9190401, Israel\\
\\
Center for Brain Science\\
 Harvard University\\
 Cambridge MA 02138, USA\\
 }

\maketitle
$^{\star}$ For correspondence: rr2980@columbia.edu 

\pagebreak{}
\begin{abstract}
Neurons and networks in the cerebral cortex must operate reliably
despite multiple sources of noise. To evaluate the impact of both
input and output noise, we determine the robustness of single-neuron
stimulus selective responses, as well as the robustness of attractor
states of networks of neurons performing memory tasks. We find that
robustness to output noise requires synaptic connections to be in
a balanced regime in which excitation and inhibition are strong and
largely cancel each other. We evaluate the conditions required for
this regime to exist and determine the properties of networks operating
within it. A plausible synaptic plasticity rule for learning that
balances weight configurations is presented. Our theory predicts an
optimal ratio of the number of excitatory and inhibitory synapses
for maximizing the encoding capacity of balanced networks for a given
statistics of afferent activations. Previous work has shown that balanced
networks amplify spatio-temporal variability and account for observed
asynchronous irregular states. Here we present a novel type of balanced
network that amplifies small changes in the impinging signals, and
emerges automatically from learning to perform neuronal and network
functions robustly.
\end{abstract}
\newpage{}

\section*{Introduction}

The response properties of neurons in many brain areas including cerebral
cortex are shaped by the balance between co-activated inhibitory and
excitatory synaptic inputs \cite{anderson_orientation_2000,wehr_balanced_2003,okun_instantaneous_2008,poo_odor_2009,atallah_instantaneous_2009}
(for a review see \cite{isaacson_how_2011}). Excitation-inhibition
balance may have different forms in different brain areas or species
and its emergence likely arise from multiple mechanisms. Theoretical
work has shown that, when externally driven, circuits of recurrently
connected excitatory and inhibitory neurons with strong synapses settle
rapidly into a state in which population activity levels ensure a
balance of excitatory and inhibitory currents \cite{van_vreeswijk_chaos_1996,vreeswijk_chaotic_1998}.
Experimental evidence in some systems indicates that synaptic plasticity
plays a role in maintaining this balance \cite{froemke_synaptic_2007,dorrn_developmental_2010,sun_fine-tuning_2010,li_broadening_2012}.
Here we address the question of what computational benefits are conferred
by the excitation-inhibition balance properties of balanced and unbalanced
neuronal circuits. Although it has been shown that networks in the
balanced states have advantages in generating fast and linear response
to changing stimuli, \cite{tsodyks_rapid_1995,van_vreeswijk_chaos_1996,vreeswijk_chaotic_1998,van_vreeswijk_course_2005},
the advantages and disadvantages of excitation-inhibition balance
for general information processing have not been elucidated (except
in special architectures; see \cite{lim_balanced_2013,boerlin_predictive_2013}).
Here we compare the computational properties of neurons operating
with and without excitation-inhibition balance and present a constructive
computational reason for strong, balanced excitation and inhibition:
it is needed for neurons to generate selective responses that are
robust to output noise, and it is crucial for the stability of memory
states in associative memory networks. The novel balanced networks
we present naturally and automatically emerge from synaptic learning
that endows neurons and networks with robust functionality. 

We begin our analysis by considering a single neuron receiving input
from a large number of afferents. We characterize its basic task as
discriminating patterns of input activation to which it should respond
by firing action potentials from other patterns which should leave
it quiescent. Neurons implement this form of response selectivity
by applying a threshold to the sum of inputs from their presynaptic
afferents. The simplest (parsimonious) model that captures these basic
elements is the binary model neuron \cite{rosenblatt_principles_1962,minsky_perceptrons:_1988},
which has been studied extensively \cite{gardner_maximum_1987,gardner_space_1988,gardner_optimal_1988,amit_interaction_1989}
and used to model a variety of neuronal circuits \cite{hopfield_neural_1982,brunel_optimal_2004,clopath_storage_2012,chapeton_efficient_2012,brunel_is_2016}.
Our work is based on including and analyzing the implications of four
fundamental neuronal features not previously considered together:
1) non-negative input, corresponding to the fact that neuronal activity
is characterized by firing rates; 2) a membrane potential threshold
for neuronal firing above the resting potential (and hense a silent
resting state); 3) sign-constrained and bounded synaptic weights,
meaning that individual synapses are either excitatory or inhibitory
and the total synaptic strength is limited; and 4) two sources of
noise, input and output noise, representing fluctuations arising from
variable stimuli and inputs and from processes within the neuron.
As will be shown, these features imply that, when the number of input
afferents is large, synaptic input must be strong and balanced if
the neuron's response selectivity is to be robust. We extend our analysis
to recurrently connected networks storing long-term memory and find
that similar balanced synaptic patterns are required for the stability
of the memory states against noise. In addition, maximizing the performance
of neurons and networks in the balanced state yields a prediction
for the optimal ratio of excitatory to inhibitory inputs in cortical
circuits.

\section*{Results}

Our model neuron is a binary unit that is either active or quiescent
depending on whether its membrane potential is above or below a firing
threshold. The potential, labeled $V_{\mathrm{PSP}}$, is a weighted
sum of inputs $x_{i}$, $i=1,2,...,N$, that represent afferent firing
rates and are thus non-negative,
\begin{equation}
V_{\mathrm{PSP}}\left(\boldsymbol{x},\ \boldsymbol{w}\right)=V_{\mathrm{rest}}+\sum_{i=1}^{N}w_{i}x_{i}\,,\label{PSP}
\end{equation}
where $V_{\mathrm{rest}}$ is the resting potential of the neuron
and $\boldsymbol{x}$ and $\boldsymbol{w}$ are $N$-component vectors
with elements $x_{i}$ and $w_{i}$ respectively. The weight $w_{i}$
represents the synaptic efficacy of the $i$'th input. If $V_{\mathrm{PSP}}\geq V_{\thMy}$
the neuron is in an active state, otherwise, it is in a quiescent
state. To implement the segregation of excitatory and inhibitory inputs,
each weight is constrained so that $w_{i}\ge0$ if input $i$ is excitatory
and $w_{i}\le0$ if input $i$ is inhibitory.

To function properly in a circuit, a neuron must respond selectively
to an appropriate set of inputs. To characterize selectivity, we define
a set of $P$ exemplar input vectors $\x^{\mu}$, with $\mu=1,2,...,P$,
and randomly assign them to two classes, denoted as `plus' and `minus'.
The neuron must respond to inputs belonging to the `plus' class by
firing (active state) and to the `minus' class by remaining quiescent.
This means that the neuron is acting as a perceptron \cite{rosenblatt_principles_1962,minsky_perceptrons:_1988,gardner_maximum_1987,gardner_space_1988,gardner_optimal_1988,amit_perceptron_1989,brunel_optimal_2004,chapeton_efficient_2012}.
We assume the $P$ input activations, $\x^{\mu}$, are drawn i.i.d.
from a distribution with non-negative means, $\bar{\boldsymbol{x}}$,
and covariance matrix, $C$ (when $N$ is large, higher moments of
the distribution of $\boldsymbol{x}$ have negligible effect). For
simplicity we assume that the stimulus average activities are the
same for all input neurons within a population, so that $\bar{x}{}_{i}=\bar{x}_{\mathrm{exc\left(inh\right)}}\geq0$,
and that $C$ is diagonal with equal variances within a population,
$\sigma_{i}^{2}=\sigma_{\mathrm{exc\left(inh\right)}}^{2}$. Note
that synaptic weights are in units of membrane potential over input
activity levels (firing rates), and hence will be measured in units
of $\left(V_{\mathrm{th}}-V_{\mathrm{rest}}\right)/\sigma_{\mathrm{exc}}$. 

We call weight vectors that correctly categorize the $P$ exemplar
input patterns, $\x^{\mu}$ for $\mu=1,2,...,P$, solutions of the
categorization task presented to the neuron. Before describing in
detail the properties of the solutions, we outline a broad distinction
between two types of possible solutions. One type is characterized
by weak synapses, \emph{i.e.}, individual synaptic weights that are
inversely proportional to the total number of synaptic inputs, $w_{i}\sim1/N$
(note that weights weaker than $\mathcal{O}\left(1/N\right)$ will
not enable the neuron to cross threshold). For this solution type,
the total excitatory and inhibitory parts of the membrane potential
are of the same order as the neuron's threshold. An alternative scenario
is a solution in which individual synaptic weights are relatively
strong, $w_{i}\sim1/\sqrt{N}$. In this case, both the total excitatory
and inhibitory parts of the potential are, individually, much greater
than the threshold, but they make approximately equal contributions,
so that excitation and inhibition tend to cancel, and the mean $V_{\mathrm{PSP}}$
is close to threshold. We call the first type of solution \emph{unbalanced}
and the second \emph{balanced}. Note that the norm of the weight vector,
$\wnorm=\sqrt{\sum_{i=1}^{N}w_{i}^{2}}$, serves to distinguish the
two types of solutions. This norm is of order of $1/\sqrt{N}$ for
unbalanced solutions and of order $1$ in the balanced case. Weights
with norms stronger than $\mathcal{O}\left(1\right)$ lead to membrane
potential values that are much larger in magnitude than the neuron's
threshold. For biological neurons postsynaptic potentials of such
magnitude can result in very high, unreasonable firing rates (although
see \cite{deneve_efficient_2016}). We therefore impose an upper bound
of the weight norm $\wnorm\le\Gamma$ where $\Gamma$ is of order
1. We now argue that the differences between unbalanced and balanced
solutions have important consequences for the way the system copes
with noise. 

As mentioned above, neurons in the central nervous system are subject
to multiple sources of noise, and their performance must be robust
to its effects. We distinguish two biologically relevant types of
noise: \emph{input noise} resulting from the fluctuations of the stimuli
and sensory processes that generate the stimulus related input $\x$;
and \emph{output noise} arising from afferents unrelated to a particular
task or from biophysical processes internal to the neuron, including
fluctuations in the effective threshold due to spiking history and
adaptation \cite{brown_muscarinic_1980,madison_control_1984,fleidervish_slow_1996}
(For theoretical modeling see \cite{benda_universal_2003}). Both
sources of noise result in trail-by-trial fluctuations of the membrane
potential $V_{\mathrm{PSP}}$ and, for a robust solution, the probability
of changing the state of the output neuron relative to the noise-free
condition must be low. The two sources of noise differ in their dependence
on the magnitude of the synaptic weights. Because input noise is filtered
through the same set of synaptic weights as the signal, its effect
on the membrane potential is sensitive to the magnitude of those weights.
Specifically, if the trial-to-trial variability of each input $x_{i}^{\mu}$
is characterized by standard deviation $\sigma_{\mathrm{in}}$, the
fluctuations it generates in the membrane potential have standard
deviation $|\w|\sigma_{\mathrm{in}}$ (Fig.~1 top row). On the other
hand, the effect of output noise is independent of the synaptic weights
$\boldsymbol{w}$. Output noise characterized by standard deviation
$\sigma_{\mathrm{out}}$ induces membrane potential fluctuations with
the same standard deviation $\sigma_{\mathrm{out}}$ for both types
of solutions (Fig.~1 bottom row). 

\begin{figure}
\includegraphics{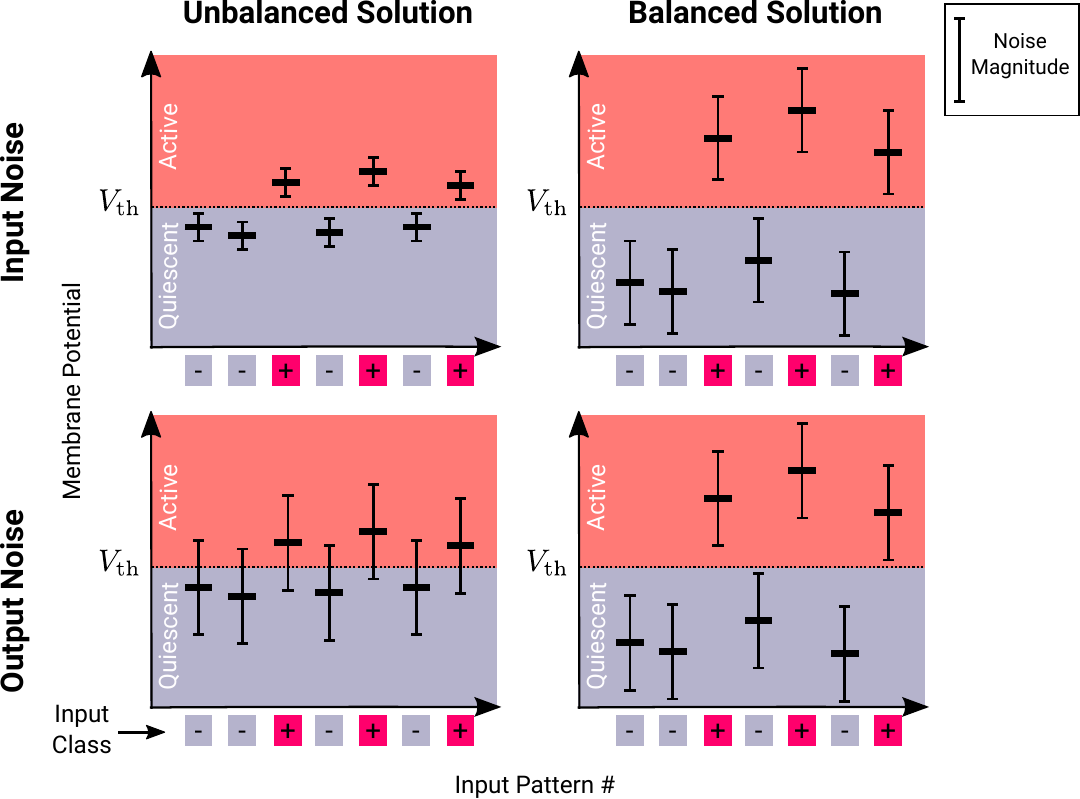}

\caption{\textbf{\small{}Only balanced solutions can be robust to both input
and output noise.}{\small{} Each panel depicts membrane potentials
resulting from different input patterns in a classification task.
Weights are unbalanced ($\protect\wnorm=\mathcal{O}\left(1/\sqrt{N}\right)$,
left column) or balanced ($\protect\wnorm=\mathcal{O}\left(1\right)$,
right column). The neuron is in an active state only if the membrane
potential is greater than the threshold $V_{\protect\thMy}$. The
input pattern class (`plus' or `minus') is specified by the squares
underneath the horizontal axis. Each input pattern determines a membrane
potential (mean, horizontal bars) that fluctuates from one presentation
to another due to input noise (top row) and output noise (bottom row).
Vertical bars depict the magnitude of the noise in each case. The
variability of the mean $V_{\mathrm{PSP}}$ across input patterns
(which is the signal differentiating input pattern classes) is proportional
to $\protect\wnorm$. As a result, the mean $V_{\mathrm{PSP}}$'s
for unbalanced solutions (left column) cluster close to the threshold
(difference from threshold $\mathcal{O}\left(1/\sqrt{N}\right)$).
For balanced solutions (right column), the mean $V_{\mathrm{PSP}}$'s
have a larger spread (potential difference $\mathcal{O}\left(1\right)$).
Input noise (fluctuations of $x_{i}$, top row) produces membrane
potential fluctuations with standard deviation that is proportional
to $|\protect\w|$, which is of $\mathcal{O}\left(1/\sqrt{N}\right)$
for unbalanced solutions (top left) and of $\mathcal{O}\left(1\right)$
for balanced solutions (top right). Output noise (bottom row) produces
membrane potential fluctuation that are independent of $\protect\wnorm$,
so it is of the same magnitude for both solution types. Thus, while
both balanced and unbalanced solutions can be robust to input noise,
only balanced solutions can also be robust to substantial output noise. }}
\end{figure}

We can now appreciate the basis for the difference in the noise robustness
of the two types of solutions. For unbalanced solutions, the difference
between the potential induced by typical `plus' and `minus' noise-free
inputs (the signal) is of the order of $\wnorm=\mathcal{O}\left(1/\sqrt{N}\right)$
(Fig.~1 left column). Although the fluctuations induced by input
noise are of this same order (Fig.~1 top left), output noise yields
fluctuations in the membrane potential of order $1,$ which is much
larger than the magnitude of the weak signal (Fig.~1 bottom left).
In contrast, for balanced solutions, the signal differentiating `plus'
and `minus' patterns is of order $\wnorm=\mathcal{O}\left(1\right)$,
which is the same order as the fluctuations induced by both types
of noise (Fig.~1 right column). Thus, we are led to the important
observation that the balanced solution provides the only hope for
producing selectivity that is robust against both types of noise.
However, there is no guaranty that robust, balanced solutions exist
or that they can be found and maintained in a manner that can be implemented
by a biological system. A key question, therefore, is under what conditions
does a balanced solution to the selectivity task exist and what are,
in detail, its robustness properties. Below, we derive conditions
for the existence of a balanced solution, analyze its properties,
and study the implications for single-neuron and network computation.
We show that, subject to a small reduction of the total information
stored in the network, robust and balanced solutions exist and can
emerge naturally when learning occurs in the presence of output noise.

\section*{Balanced and unbalanced solutions}

We begin by presenting the results of an analytic approach \cite{gardner_maximum_1987,gardner_optimal_1988,gardner_space_1988}
for determining existence conditions and analyzing properties of weights
that generate a specified selectivity,\textbf{ }independent of the
particular method or learning algorithm used to find the weights (\emph{SI
Methods}). We validate the theoretical results by using numerical
methods that can determine the existence of such weights and find
them if they exist (\emph{Methods}). 

\begin{figure*}
\includegraphics{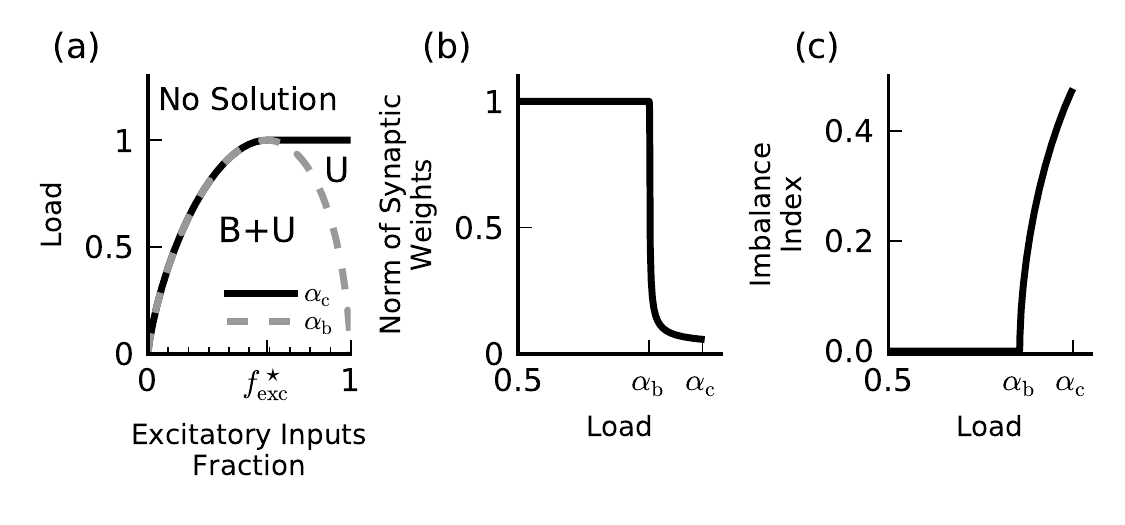} 

\caption{\label{fig:res_2} \textbf{\small{}Balanced and unbalanced solutions.}{\small{}
(a) Perceptron solutions as a function of load and fraction of excitatory
weights. Above the capacity line ($\alpha_{\mathrm{c}}\left(f_{\mathrm{exc}}\right)$,
black line) no solution exists. Balanced solutions exist only below
the balanced capacity line, ($\alpha_{\mathrm{b}}\left(f_{\mathrm{exc}}\right)$,
dashed gray line). Between the balanced capacity and maximum capacity
lines, only unbalanced solutions exist (U). On the other hand, below
the balanced capacity line, unbalanced solutions coexist with balanced
ones (B+U). (b) The norm of the synaptic weight vector of typical
solutions as a function the load (in units of }$\left(V_{\mathrm{th}}-V_{\mathrm{rest}}\right)/\sigma_{\mathrm{exc}}${\small{}).
Below $\alpha_{\mathrm{b}}$ the norm is clipped at its upper bound
$\Gamma$ (in this case $\Gamma=1$). Above $\alpha_{\mathrm{b}}$
the norm collapses and is of order $1/\sqrt{N}$ (shown here for $N=3000$).
(c) The input imbalance index ($\text{IB}$, eq. \ref{IB}) of typical
solutions as a function of the load. Note the sharp onset of imbalance
above $\alpha_{\mathrm{b}}$. In (b) and (c) $f_{\mathrm{exc}}=0.8$,
yielding $\alpha_{\mathrm{c}}=1$. See }\emph{\small{}Methods}{\small{}
for other parameters used. For simulation results see Fig. S1. }}
\end{figure*}

When the number of patterns $P$ is too large, solutions may not exist.
The maximal value of $P$ that permits solutions is proportional to
the number of synapses, $N$, so a useful measure is the ratio $\alpha=P/N$,
which we call the load. The capacity, denoted as $\alpha_{\mathrm{c}}$,
is the maximal load that permits solutions to the task. The capacity
depends on the relative number of `plus' and `minus' input patterns.
For simplicity we assume throughout that the two classes are equal
in size (but see \emph{SI Methods}). A classic result for the perceptron
with weights that are not sign constrained is that the capacity is
$\alpha_{\mathrm{c}}=2$ \cite{cover_geometrical_1965,venkatesh_epsilon_1986,gardner_maximum_1987}.
For the `constrained perceptron' considered here, we find that $\alpha_{\mathrm{c}}$
depends also on the fraction of excitatory afferents, denoted by $f_{\mathrm{exc}}$.
This fraction is an important architectural feature of neuronal circuits
and varies in different brain systems. For $f_{\mathrm{exc}}=0$,
namely purely inhibitory circuit, the capacity vanishes, because when
all the input to the neuron is inhibitory, $V_{\mathrm{PSP}}$ cannot
reach threshold and the neuron is quiescent for all stimuli. When
the circuit includes excitatory synapses, the task can be solved by
appropriate shaping of the strength of the excitatory and inhibitory
synapses, and this ability increases the larger the fraction of excitatory
synapses is. Therefore, For $f_{\mathrm{exc}}>0$, $\alpha_{\mathrm{c}}$
increases with $f_{\mathrm{exc}}$ up to a maximum of $\alpha_{\mathrm{c}}=1$
(half the capacity of an unconstrained perceptron) for fractions equal
or greater than a critical fraction $f_{\mathrm{exc}}=f_{\mathrm{exc}}^{\star}$.
This dependence can be summarized by the capacity curve $\alpha_{\mathrm{c}}\left(f_{\mathrm{exc}}\right)$
(Fig. \ref{fig:res_2}a, black line) bounding the range of loads which
admit solutions for the different excitatory/inhibitory ratios. 

Interestingly, $f_{\mathrm{exc}}^{\star}$ depends on the statistics
of the inputs (\emph{SI Methods}). We denote the coefficient of variation
(CV) of the excitatory and inhibitory input activities by $\mathrm{CV_{exc}}=\sigma_{\mathrm{exc}}/\bar{x}_{\mathrm{exc}}$
and $\mathrm{CV_{inh}}=\sigma_{\mathrm{inh}}/\bar{x}_{\mathrm{inh}}$
, respectively. These, measure the degree of stimulus tuning of the
two afferent populations. In terms of these quantities, the critical
excitatory fraction is

\begin{equation}
f_{\mathrm{exc}}^{\star}=\frac{\mathrm{CV_{exc}}}{\mathrm{CV_{exc}}+\mathrm{CV_{inh}}}\ .\label{f_star}
\end{equation}
In other words, the critical ratio between the number of excitatory
and inhibitory afferents ($f_{\mathrm{exc}}^{\star}$/(1-$f_{\mathrm{exc}}^{\star}$))
equals the ratio of their degree of tuning. To understand the origin
of this result, we note that to maximize the encoding capacity, the
relative strength of the weights should be inversely proportional
to the standard deviation of their afferents, $\bar{w}_{\mathrm{exc\left(inh\right)}}\propto1/\sigma_{\mathrm{exc\left(inh\right)}}$,
implying that the mean total synaptic inputs is proportional to $f_{\mathrm{exc}}\bar{w}_{\mathrm{exc}}\bar{x}_{\mathrm{exc}}+f_{\mathrm{inh}}\bar{w}_{\mathrm{inh}}\bar{x}_{\mathrm{inh}}=f_{\mathrm{exc}}/\mathrm{CV}_{\mathrm{exc}}-f_{\mathrm{inh}}/\mathrm{CV}\mathrm{_{inh}}$
where $f_{\mathrm{inh}}=1-f_{\mathrm{exc}}$. For excitatory fraction
$f_{\mathrm{exc}}>f_{\mathrm{exc}}^{\star}$ this mean total synaptic
inputs is positive, allowing the voltage to reach the threshold and
the neuron to implement the required selectivity task with optimally
scaled weights. Thus, the capacity of the neuron is unaffected by
changes in $f_{\mathrm{exc}}$ in the range $f_{\mathrm{exc}}^{\star}\le f_{\mathrm{exc}}\le1$.
For excitatory fraction $f_{\mathrm{exc}}<f_{\mathrm{exc}}^{\star}$
the neuron cannot remain responsive (reach threshold) with optimally
scaled weights, and thus the capacity is reduced. 

In cortical circuits, inhibitory neurons tend to fire at higher firing
rates and are thought to be more broadly tuned than excitatory neurons
\cite{poo_odor_2009,liu_visual_2009,kerlin_broadly_2010}, implying
$f_{\mathrm{exc}}^{\star}>0.5$ (\emph{SI Methods}). This is consistent
with the abundance of excitatory synapses in cortex. However, input
statistics that make $f_{\mathrm{exc}}^{\star}<0.5$ do not change
the qualitative behavior we discuss (\emph{SI Methods} and Fig. S2a).

For load levels below the capacity, many synaptic weight vectors solve
the selectivity task and we now describe the properties of the different
solutions. In particular, we investigate the parameter regimes where
balanced or unbalanced solutions exist. We find that unbalanced solutions
with weights vector norm of order $1/\sqrt{N}$ exist for all load
values below $\alpha_{\mathrm{c}}$. As for the balanced solutions
with weight vector norms of order $1$, they exist below a critical
value $\alpha_{\mathrm{b}}$ which may be smaller than $\alpha_{\mathrm{c}}$
. Specifically, for $f_{\mathrm{exc}}\le f_{\mathrm{exc}}^{\star}$
balanced solutions exist for all load values below capacity, \emph{i.e}.,
$\alpha_{\mathrm{b}}=\alpha_{\mathrm{c}}$. For $f_{\mathrm{exc}}>f_{\mathrm{exc}}^{\star}$,
$\alpha_{\mathrm{b}}$ is smaller than $\alpha_{\mathrm{c}}$ and
decreases with $f_{\mathrm{exc}}$ until it vanishes at $f_{\mathrm{exc}}=1$
(Fig.~\ref{fig:res_2}a, dashed gray line). The absence of balanced
solutions for $f_{\mathrm{exc}}=1$ is clear, as there is no inhibition
to balance the excitatory inputs. Furthermore, the synaptic excitatory
weights must be weak (scaling as $1/N$ ) to ensure that $V_{\mathrm{PSP}}$
remains close to threshold (slightly above it for `plus' patterns
and slightly below it for `minus' ones). For $1\geq f_{\mathrm{exc}}>f_{\mathrm{exc}}^{\star}$
the predominance of excitatory afferents precludes a balanced solution
if the load is high, i.e., $\alpha_{\mathrm{b}}\leq\alpha\leq\alpha_{\mathrm{c}}$.
As argued above and shown below, the balanced solution is more robust
than the unbalanced solution. Hence, we can identify $f_{\mathrm{exc}}^{\star}$
as the optimal fraction of excitatory input, because it is the fraction
of excitatory afferents for which the capacity of \emph{balanced}
solutions is maximal. 

For loads below $\alpha_{\mathrm{b}}$ both balanced and unbalanced
solutions exist, raising the question what would be the character
of a weight vector that is sampled randomly from the space of all
possible solutions. Our theory predicts that whenever the balance
solution exists, the vast majority of the solutions are balanced and
furthermore have a weight vector norm that is saturated at the upper
bound $\Gamma$. Thus, for $f_{\mathrm{exc}}>f_{\mathrm{exc}}^{\star}$,
the \emph{typical} solution undergoes a transition from balanced to
unbalanced weights as $\alpha$ crosses the balanced capacity line
$\alpha_{\mathrm{b}}\left(f_{\mathrm{exc}}\right)$. At this point
the norm of the solution collapses from $\Gamma$ to $\wnorm\sim1/\sqrt{N}$
(Fig. \ref{fig:res_2}b). 

As explained above, for balanced solutions we expect to find a near
cancellation of the total excitatory and inhibitory inputs. Our theory
confirms this expectation. To measure the degree of E-I cancellation
for any solution, we introduce the \emph{imbalance index},

\begin{equation}
\text{{IB}}=\frac{\sum_{i}w_{i}\bar{x}_{i}}{\sum_{i\in\mathrm{exc}}w_{i}\bar{x}_{i}-\sum_{i\in\mathrm{inh}}w_{i}\bar{x}_{i}}\ ,\,\label{IB}
\end{equation}
where the bar symbol denotes an average over all the input patterns
($\mu$). Whereas for the unbalanced solution the IB is of order $1$,
for the balanced solution it is small, of order $1/\sqrt{N}$. Thus,
the typical solution below $\alpha_{\mathrm{b}}$ has zero imbalance
(to leading order in $N)$, but the imbalance increases sharply as
$\alpha$ increases beyond $\alpha_{\mathrm{b}}$ (Fig. \ref{fig:res_2}c). 

\subsection*{Noise robustness of balanced and unbalanced solutions}

To characterize the effect of noise on the different solutions, we
introduce two measures: input-robustness $\kappa_{\mathrm{in}}$ and
output robustness $\kappa_{\mathrm{out}}$, which characterize the
robustness of the noise-free solutions to the addition of two types
of noise. To ensure robustness to \emph{output} noise, the noise-free
membrane potential that is the closest to the threshold must be sufficiently
far from it. Thus we define

\begin{equation}
\kappa_{\mathrm{out}}=\min_{\mu}\left|\sum_{i=1}^{N}w_{i}x_{i}^{\mu}-1\right|\:,
\end{equation}
where the minimum is taken over all the input patterns in the task
and the threshold is 1 (because we measure the weights in units of
$\left(V_{\mathrm{th}}-V_{\mathrm{rest}}\right)/\sigma_{\mathrm{exc}}$).
The second measure, which characterizes robustness to\emph{ input
}noise, must take into account the fact that the fluctuations in the
membrane potential induced by this form of noise scale with the size
of the synaptic weights. Hence, $\kappa_{\mathrm{in}}=\kappa_{\mathrm{out}}/\wnorm$
($\kappa_{\mathrm{in}}$ corresponds to the notion of \emph{margin}
in machine learning \cite{vapnik_nature_2000}). Efficient algorithms
for finding the solution with maximum $\kappa_{\mathrm{in}}$ have
been studied extensively \cite{vapnik_nature_2000,bottou_support_2007}.
We have developed a novel efficient algorithm for finding solutions
with maximum $\kappa_{\mathrm{out}}$ (\emph{SI Methods}). 

We now ask what are the possible values of the input and output robustness
of unbalanced and balanced solutions. Our theory predicts that the
majority of both balanced and unbalanced solutions have vanishingly
small values of $\kappa_{\mathrm{in}}$ and $\kappa_{\mathrm{out}}$
and are thus very sensitive to noise. However, for a given load (below
capacity) robust solutions do exist, with a spectrum of robustness
values up to maximal values, $\kappa_{\mathrm{in}}^{\mathrm{max}}>0$
and $\kappa_{\mathrm{out}}^{\mathrm{max}}>0$. Since the magnitude
of $\boldsymbol{w}$ scales both signal and noise in the inputs, $\kappa_{\mathrm{in}}^{\mathrm{max}}$
is not sensitive to $\wnorm$ and hence is of $\mathcal{O}\left(1\right)$
for both unbalanced and balanced solutions. On the other hand, $\kappa_{\mathrm{out}}^{\max}=\kappa_{\mathrm{in}}^{\mathrm{max}}\wnorm$
is proportional to $\wnorm$. Thus, we expect $\kappa_{\mathrm{out}}^{\max}$
to be of $\mathcal{O}\left(1\right)$ when balanced solutions exist
and of $\mathcal{O}\left(1/\sqrt{N}\right)$ when only unbalanced
solutions exist. In addition, we expect that increasing the load will
reduce the value of $\kappa_{\mathrm{in}}^{\mathrm{max}}$ and $\kappa_{\mathrm{out}}^{\mathrm{max}}$
as the number of constraints that need to be satisfied by the synaptic
weights increases. 

\begin{figure}
\includegraphics{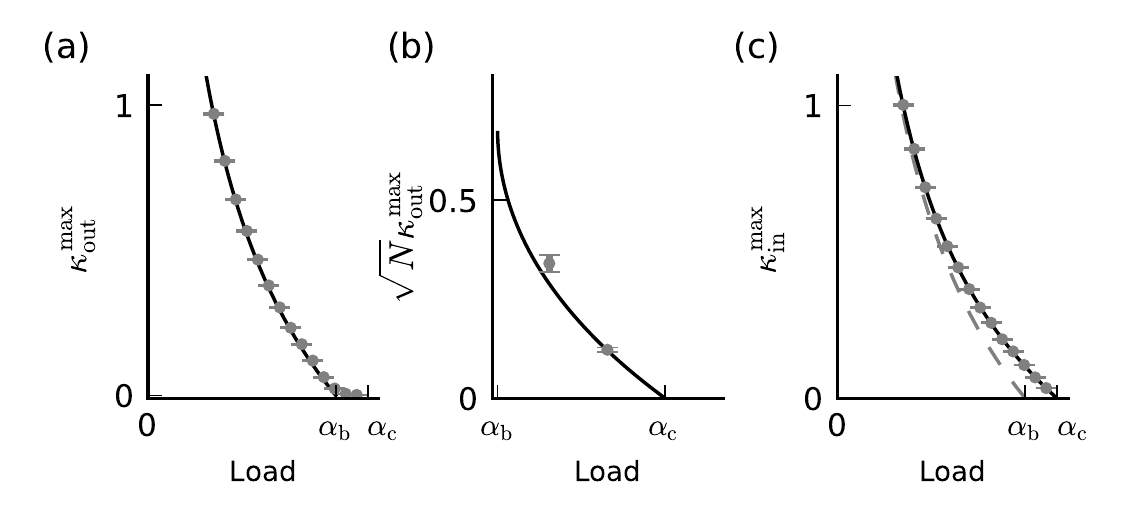} \caption{\label{fig:res_1} \textbf{\small{}Maximal values of input and output
robustness.}{\small{} (a) Maximal value of $\kappa_{\mathrm{out}}$
vs. load (in units of $\Gamma\sigma_{\mathrm{exc}}/\left(V_{\mathrm{th}}-V_{\mathrm{rest}}\right)$).
No solutions exist above the maximal $\kappa_{\mathrm{out}}$ line
($\kappa_{\mathrm{out}}^{\mathrm{max}},$ black). Below $\kappa_{\mathrm{out}}^{\mathrm{max}}$,
for output robustness that is of order 1, only balanced solutions
exist. (b) Maximal value of $\kappa_{\mathrm{out}}$ for loads between
$\alpha_{\mathrm{b}}$ and $\alpha_{\mathrm{c}}$ (in units of $\sigma_{\mathrm{exc}}/\bar{x}_{\mathrm{exc}}$).
In this range only unbalanced solution exist and the maximal $\kappa_{\mathrm{out}}$
values (black line) scale as $1/\sqrt{N}$. (c) Maximal value of $\kappa_{\mathrm{in}}$
vs. load (in units of $\sigma_{\mathrm{exc}}$). No solutions exist
above the maximal $\kappa_{\mathrm{in}}$ line ($\kappa_{\mathrm{in}}^{\mathrm{max}}$,
black). For the parameters used, solutions that achieve $\kappa_{\mathrm{in}}^{\mathrm{max}}$
are unbalanced. The maximal value of $\kappa_{\mathrm{in}}$ for balanced
solutions (dashed gray line) is not far from the $\kappa_{\mathrm{in}}^{\mathrm{max}}$
and is attained by solutions that maximize $\kappa_{\mathrm{out}}$
for $\alpha<\alpha_{\mathrm{b}}$. In all panels, theory and numerical
results are depicted in black or gray lines and gray dots respectively.
Error-bars depict standard error of the mean. See }\emph{\small{}Methods}{\small{}
for parameters used. For further simulation results see Fig. S3.}}
\end{figure}

In Fig. \ref{fig:res_1} we present the values of $\kappa_{\mathrm{in}}^{\mathrm{max}}$
and $\kappa_{\mathrm{out}}^{\mathrm{max}}$ vs. the load. As expected,
we find that the value of both $\kappa_{\mathrm{in}}^{\mathrm{max}}$
and $\kappa_{\mathrm{out}}^{\mathrm{max}}$ reach zero as the load
approaches the capacity, $\alpha_{\mathrm{c}}$ (and diverges, as
$N\rightarrow\infty$, for vanishingly small loads). However $\kappa_{\mathrm{out}}^{\mathrm{max}}$
is only substantial (of order 1) and proportional to $\Gamma$ below
$\alpha_{\mathrm{b}}$ where balanced solutions exist (Fig. \ref{fig:res_1}a-b).
In contrast $\kappa_{\mathrm{in}}^{\mathrm{max}}$ remains of order
1 up to the full capacity, $\alpha_{\mathrm{c}}$ (Fig. \ref{fig:res_1}c).
What are the properties of `optimal' solutions that achieve the maximal
robustness to either input or output noise? We find that the solutions
that achieve the maximal output robustness, $\kappa_{\mathrm{out}}^{\mathrm{max}}$,
are balanced for all $\alpha\le\text{\ensuremath{\alpha}}_{\mathrm{b}}$
and their norm saturates the upper bound, $\Gamma$ (Fig. S3b). Interestingly,
for a wide range of input parameters (\emph{SI Methods}, and Fig.
S2b), solutions that achieve the maximal input robustness, $\kappa_{\mathrm{in}}^{\mathrm{max}}$,
are unbalanced solutions (Fig. S3c). Nevertheless, we find that below
the critical balance load, $\alpha_{\mathrm{b}}$, the $\kappa_{\mathrm{in}}$
values of the balanced maximal $\kappa_{\mathrm{out}}$ solutions
are of the same order as, and indeed close to, $\kappa_{\mathrm{in}}^{\mathrm{max}}$
(Fig. \ref{fig:res_1}c, dashed gray line). In fact, the balanced
solution with maximal $\kappa_{\mathrm{out}}$ also posses the maximal
value of $\kappa_{\mathrm{in}}$ that is possible for balanced solutions.

We conclude that solutions that are robust to both input and output
noise exist for loads less than $\alpha_{\mathrm{b}}$ which for $f_{\mathrm{exc}}>f_{\mathrm{exc}}^{\star}$
is smaller than $\alpha_{\mathrm{c}}$ However, as long as $f_{\mathrm{exc}}$
is close to $f_{\mathrm{exc}}^{\star}$, the reduction in capacity
from $\alpha_{\mathrm{c}}$ to $\alpha_{\mathrm{b}}$ imposed by the
requirement of robustness is small. 

\subsection*{Balanced and unbalanced solutions for spiking neurons}

Neurons typically receive their input and communicate their output
through action potentials. Thus, a fundamental question is how will
the introduction of spike-based input and spiking output affect our
results. Here we show that the main properties of balanced and unbalanced
synaptic efficacies, as discussed above, remain when the inputs are
spike trains and the model neuron implements spiking and membrane
potential reset mechanisms.

\begin{figure*}
\includegraphics{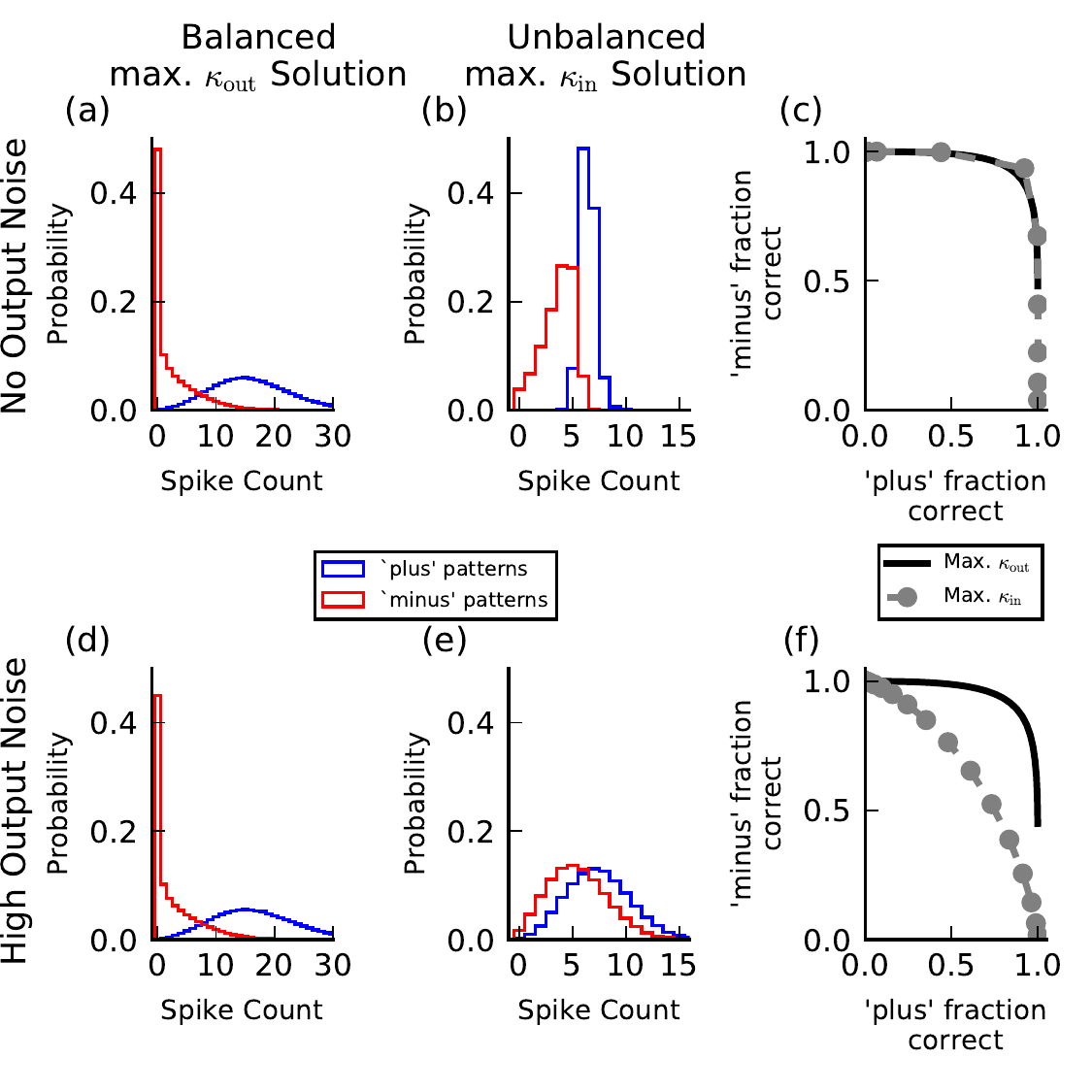}

\caption{\label{fig:spikes}\textbf{\small{}Selectivity in a spiking model.}{\small{}
Panels (a)-(b) (panels (d)-(e)) depict the output of an LIF neuron
with no (high) output noise for the balanced maximal $\kappa_{\mathrm{out}}$
solution ((a) and (d)) and the unbalanced maximal $\kappa_{\mathrm{in}}$
solution ((b) and (e)). Panels (c) and (f) depict the ROC curves for
the two solutions under the no output noise (in (c)) and high output
noise (in (f)) conditions obtained as the decision threshold ($n_{\mathrm{thr}}$)
is modified from 0 to $\infty$. Consistently with the results of
the perceptron, the performance of the two solutions with no output
noise is very similar with a slight advantage for the maximal $\kappa_{\mathrm{in}}$
solution. With higher levels of output noise, the performance of the
unbalanced maximal $\kappa_{\mathrm{in}}$ solution quickly deteriorates,
whereas the performance of the balanced maximal $\kappa_{\mathrm{out}}$
solution is only slightly affected. $\protect\wnorm$ of the balanced
solution was chosen to equalize the mean output spike count across
all patterns in both solutions (mean $n_{\mathrm{spike}}\sim4$).
See }\emph{\small{}Methods}{\small{} for parameters used.}}
\end{figure*}

We consider a leaky integrate-and-fire (LIF) neuron that is required
to perform the same binary classification task we considered using
the perceptron. Each input is characterized by a vector of firing
rates, $\boldsymbol{x}^{\mu}$. Each afferent generates a Poisson
spike train over an interval from time $t=0$ to $t=T$, with mean
rate $r_{i}\propto x_{i}^{\mu}$. The LIF neuron integrates these
input spikes (\emph{Methods}), and emits an output spike whenever
its membrane potential crosses a firing threshold. After each output
spike, the membrane potential is reset to the resting potential, and
the integration of inputs continues. We define the output state of
the LIF neuron using the total number of output spikes $n_{\mathrm{spikes}}$:
the neuron is quiescent if $n_{\mathrm{spikes}}\le n_{\mathrm{thr}}$
and active if $n_{\mathrm{spikes}}>n_{\mathrm{thr}}$ were $n_{\mathrm{thr}}$
is chosen to maximize classification performance. We will not discuss
the properties of learning in LIF neurons \cite{gutig_tempotron:_2006,memmesheimer_learning_2014,gutig_spiking_2016,rubin_neural_2013,gutig_spike_2014},
but instead test the properties of the solutions (weights) obtained
from the perceptron model when they are used for the LIF neuron. In
particular, we compare the performance of the balanced, maximal $\kappa_{\mathrm{out}}$
solution and the unbalanced, maximal $\kappa_{\mathrm{in}}$ solution.
When the synaptic weight of the LIF neuron are set according to the
two perceptron solutions, the mean output of the LIF neuron correctly
classifies the input patterns (according to the desired classification;
Fig. S4). Consistently with the results for the perceptron, we find
that with no output noise the performance of both solutions is good,
even in the presence of the substantial input noise caused by Poisson
fluctuations in the number of input spikes and their timings (Fig.
\ref{fig:spikes}a-c). When the output noise magnitude is increased
(\emph{Methods}), however, the performance of the unbalanced maximal
$\kappa_{\mathrm{in}}$ solution quickly deteriorates, whereas the
performance of the balanced maximal $\kappa_{\mathrm{out}}$ solution
remains largely unaffected (Fig. \ref{fig:spikes}d-f). Thus, the
spiking model recapitulates the general results found for the perceptron.

\subsection*{Balanced and unbalanced synaptic weights in associative memory networks }

Thus far, we have considered the selectivity of a single neuron, but
our results also have important implications for recurrently connected
neuronal networks, in particular recurrent networks implementing associative
memory functions. Models of associative memory in which stable fixed
points of the network dynamics represent memories, and memory retrieval
corresponds to the dynamic transformation of an initial state to one
of the memory-representing fixed points, have been a major focus of
memory research for many years \cite{hopfield_neural_1982,amit_storing_1985,tsodyks_enhanced_1988,roudi_balanced_2007,chapeton_efficient_2012,brunel_is_2016}.
For the network to function as an associative memory, memory states
must have large basins of attraction so that the network can perform
pattern completion, recalling a memory from an initial state that
is similar but not identical to it. In addition, memory retrieval
must be robust to output noise. As we will show, the variables $\kappa_{\mathrm{in}}$
and $\kappa_{\mathrm{out}}$ for the synaptic weights projecting onto
individual neurons in the network are closely related to the sizes
of the basins of attraction of the memories and the robustness to
output noise, respectively.

\begin{figure}
\begin{minipage}[t]{1\columnwidth}%
\begin{wrapfigure}[34]{o}{0.55\columnwidth}%
\includegraphics{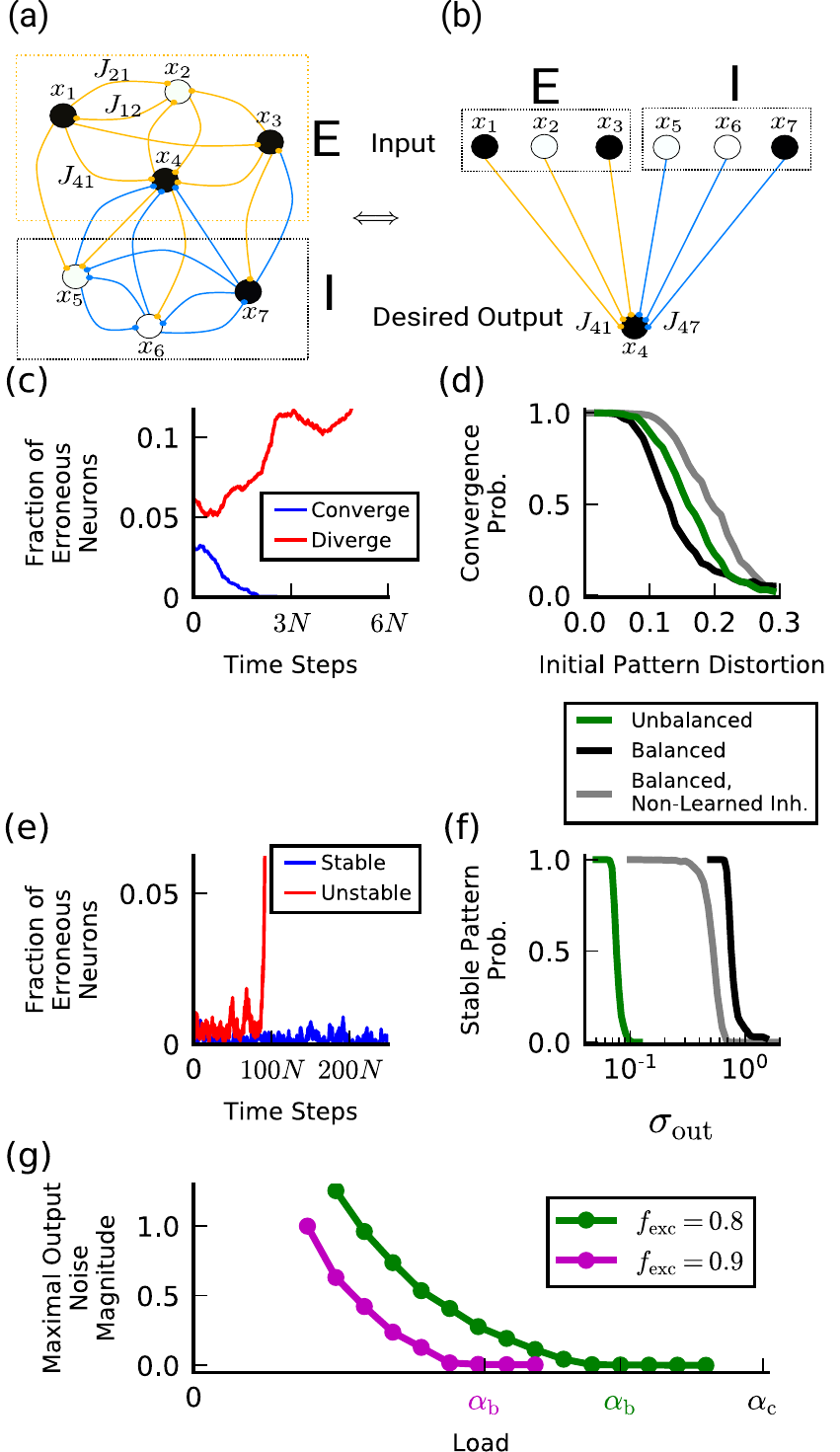} \end{wrapfigure}%
\noindent\refstepcounter{figure}Figure \thefigure:\label{fig:res_3}\textbf{
}\textbf{\small{}Recurrent associative memory network constructed
using single-neuron feedforward learning.}{\small{} (a) A fully connected
recurrent network of excitatory (E) and inhibitory (I) neurons in
a particular memory state. Active (quiescent) neurons are shown in
black (white). Excitatory and inhibitory synaptic connections ($J_{ij}$)
are shown in yellow and blue, respectively (not all connections are
depicted). Lines symbolize axons, and synapses are shown as small
circles. (b) To find an appropriate $J_{ij}$, the postsynaptic weights
of each neuron are set using the memory-state activities of the other
neurons as input and its own memory state as the desired output. In
this example, neuron \#4 will implement its desire memory state through
modification of the weights $J_{4j}$ for $j=1,2,3,5,6,7$. (c) and
(e) show the fraction of erroneous (different from a given memory
pattern) neurons in the network as a function of time. (c) Network
dynamics with $\sigma_{\mathrm{out}}=0$. An initial state of the
network can either converge to the memory state (blue) or diverge
to other network states (red). (d) Probability of converging to a
memory state steps vs.\ initial pattern distortion (}\emph{\small{}Methods}{\small{})
for a network with unbalanced maximal $\kappa_{\mathrm{in}}$ weights
(green), a network with balanced maximal $\kappa_{\mathrm{out}}$
weights (black) and a network with balanced maximal $\kappa_{\mathrm{out}}$
weights with unlearned inhibition (gray, see text). (e) Network dynamics
with $\sigma_{\mathrm{out}}>0$. The network is initialized at the
memory state. The dynamics can be stable (blue; the network remains
close to the memory state), or unstable (red; the network diverge
to another state). (f) Probability of stable dynamics for at least
$500N$ time steps for networks initialized at the memory state in
the presence of output noise vs.\ $\sigma_{\mathrm{out}}$. Colors
are the same as in (d). (g) Maximal output noise magnitude vs.\ load
for networks with balanced synaptic weights matrix maximizing $\kappa_{\mathrm{out}}$.
Similarly to $\kappa_{\mathrm{out}}$, the maximal output noise magnitude
is of order 1 only below $\alpha_{\mathrm{b}}$. Above it, even though
solutions exists they are extremely sensitive to output noise. Results
are shown for $f_{\mathrm{exc}}=0.8$ (green) and $f_{\mathrm{exc}}=0.9$
(magenta). See }\emph{\small{}Methods}{\small{} for parameters used.}%
\end{minipage}

\end{figure}

We consider a network that consists of $Nf_{\mathrm{exc}}$ excitatory
and $N\left(1-f_{\mathrm{exc}}\right)$ inhibitory, recurrently connected
binary neurons. The network operates in discrete time steps and, at
each step the state of one randomly chosen neuron, $i$, is updated
according to 
\begin{equation}
s_{i}\left(t+1\right)=\Theta\left[\sum_{j\ne i}J_{ij}s_{j}\left(t\right)+\eta_{\mathrm{out}}\left(t\right)-1\right]\ .
\end{equation}
Here $\Theta\left(x\right)=1$ for $x\geq0$ and 0 otherwise, $J_{ij}$
is the weight of the synapse from neuron $j$ to neuron $i$, and
$\eta_{\mathrm{out}}(t)$, the output noise, is a Gaussian random
variable with standard deviation $\sigma_{\mathrm{out}}$. $P$ randomly
chosen binary activity patterns $\left\{ \mathbf{s}^{\mu}\right\} ,\ \mu=1,2,...,P$
(where each $s_{i}^{\mu}=\left\{ 0,1\right\} $) representing the
stored memories are encoded in the recurrent synaptic matrix $J$
such that they will be fixed points of the network dynamics. This
is achieved by treating each neuron, say $i,$ as a perceptron with
a weight vector $\boldsymbol{w}^{i}=\{J_{ij}\}_{j\neq i}$ that maps
its inputs $\{s_{j}^{\mu}$\} from all other neurons to its desired
output $s_{i}^{\mu}$ for each memory state (Fig. \ref{fig:res_3}a-b,
\emph{Methods}). This creates an attractor network in which the memory
states are stable fixed points of the dynamics in the noise-free condition
($\sigma_{\mathrm{out}}=0$) \cite{gardner_maximum_1987}. 

The capacity of the memory network is defined as the maximal load
for which the memory patterns are stable fixed points of the dynamics.
The capacity of a single neuron perceptron depends on the statistics
of its desired output (which in our case is the sparsity of activity
across memory states). Since this statistic may be different in excitatory
and inhibitory populations, the single neuron capacity of the two
populations may vary, hence the global capacity of the recurrent network
is the minimum of the single-neuron capacities of the two neuron types.
As long as $P$ is smaller than this critical capacity, a recurrent
weight matrix exists for which all $P$ memory states are stable fixed
points of the noiseless dynamics. However such solutions are not unique,
and the choice of a particular matrix can endow the network with different
robustness properties. As stated above, to properly function as an
associative memory the fixed points have large basins of attraction.
Corruption of the initial state away from the parent memory pattern
introduces variability into the inputs of each neuron for subsequent
dynamic iterations and hence is equivalent to injecting input noise
in the single-neuron feedforward case. Therefore a large basin of
attraction is achieved when the matrix $J$ yields a large input noise
robustness for each neuron in the (noise free) fixed points \cite{krauth_basins_1988,krauth_roles_1988}.
The requirement that the memory states and retrieval will be robust
against output noise is satisfied when $J$ yields a large output
noise robustness for each neuron in the (noise free) fixed points.
We therefore consider two types of recurrent connections: one in which
each row of $J$ is a weight vector that maximizes $\kappa_{\mathrm{in}}$
and hence, in the chosen parameter regime, is necessarily unbalanced;
and a second in which the rows of the connection matrix correspond
to balanced solutions that maximize $\kappa_{\mathrm{out}}$. 

We estimate the basins of attraction of the memory patterns numerically
by initializing the network in states that are corrupted versions
of the memory states (\emph{Methods}) and observing if the network,
with $\mathrm{\sigma_{out}}=0$, converges to the parent memory state
(Fig.~\ref{fig:res_3}c, blue) or diverges away from it (Fig.~\ref{fig:res_3}c,
red). We define the size of the basin of attraction as the maximum
distortion in the initial state that assures convergence to the parent
memory with high probability. 

Comparing the basins of attraction of the two types of networks, we
find that the mean basin of attraction of the unbalanced network is
moderately larger than that of the balanced one (Fig.~\ref{fig:res_3}d),
consistent with the slightly lower value of $\kappa_{\mathrm{in}}$
in the balanced case (Fig.~\ref{fig:res_3}d).  On the other hand,
the behavior of the two networks is strikingly different in the presence
of output noise. To illustrate this, we start each network at a memory
state and determine if it is stable, (remains in the vicinity of this
state for an extended period of time) despite the noise in the dynamics
(Fig.~\ref{fig:res_3}e). We estimate the output noise tolerance
of the network by measuring the maximal value of $\sigma_{\mathrm{out}}$
for which the memory states are stable (Fig. \ref{fig:res_3}f). We
find that memory states in the balanced solution with maximal $\kappa_{\mathrm{out}}$
are stable for noise levels that (for the network sizes used in the
simulation) are an order of magnitude larger than for the unbalanced
network with maximal $\kappa_{\mathrm{in}}$ (Fig.~\ref{fig:res_3}f).

Finally, we ask how the noise robustness of the memory states in the
balanced network depends on the number of memories. As shown in Fig.
\ref{fig:res_3}f, for a fixed level of load below capacity, memory
patterns are stable ($P_{\mathrm{stable}}>0.5$) as long as levels
of noise remain below a threshold value, which we denote as $\sigma_{\mathrm{out}}^{\mathrm{max}}\left(\alpha\right)$.
When $\sigma_{\mathrm{out}}$ increases beyond $\sigma_{\mathrm{out}}^{\mathrm{max}}\left(\alpha\right)$
stability of the memory states rapidly deteriorates. The critical
noise function $\sigma_{\mathrm{out}}^{\mathrm{max}}\left(\alpha\right)$
decreases smoothly from a large value at small $\alpha$ to zero at
a level of load, $\alpha_{\mathrm{b}}$. This load coincides with
the maximal load for which both excitatory and inhibitory neurons
have balanced solution (Fig. \ref{fig:res_3}g). For loads $\alpha_{\mathrm{b}}<\alpha<\alpha_{\mathrm{c}}$,
all solutions are unbalanced, hence the magnitude of the stochastic
dynamical component can be at most of order $1/\sqrt{N}$. 

\subsection*{The role of inhibition in associative memory networks}

In our associative memory network model, we assumed that both excitatory
and inhibitory neurons code desired memory states and that all network
connections are modified by learning. Most previous models of associative
memory that separate excitation and inhibition assume that memory
patterns are restricted to the excitatory population, whereas inhibition
provides stabilizing inputs \cite{amit_associative_1989,golomb_willshaw_1990,hasselmo_acetylcholine_1993,barkai_modulation_1994,van_vreeswijk_course_2005,roudi_balanced_2007}.
To address the emergence of balanced solution in scenarios where the
inhibitory neurons do not represent long-term memories, we studied
an architecture where I to E, I to I and E to I connections are random
sparse matrices with large amplitudes, resulting in inhibitory activity
patterns driven by the excitatory memory states. In such conditions,
the inhibitory subnetwork exhibits irregular asynchronous activity
with an overall mean activity that is proportional to the mean activity
of the driving excitatory population \cite{van_vreeswijk_chaos_1996,kadmon_transition_2015,harish_asynchronous_2015}.
Although the mean inhibitory feedback provided to the excitatory neurons
can balance the mean excitation, the variability in this feedback
injects substantial noise onto the excitatory neurons, which degrades
system performance (\emph{SI Methods}). This variability stems from
the differences in inhibitory activity patterns generated by the different
excitatory memory states (albeit with the same mean). Additional noise
is caused by the temporal irregular activity of the chaotic inhibitory
dynamics. Next we ask whether system's performance can be improved
through plasticity in the I to E connections for which some experimental
evidence exist \cite{nugent_ltp_2008,chevaleyre_heterosynaptic_2003,amit_interaction_1989,damour_inhibitory_2015,mcbain_presynaptic_2009}.
Indeed, we find an appropriate plasticity rule for this pathway (\emph{SI
Methods}) that suppresses the spatio-temporal fluctuations in the
inhibitory feedback, yielding a balanced state that behaves similarly
to the fully learned networks described above (Fig. \ref{fig:res_3}d,
\ref{fig:res_3}f, gray lines). Interestingly, in this case the basins
of attraction of the balanced network are comparable to or even larger
than the basins of the unbalanced fully learned network (compared
gray to green curves in Fig. \ref{fig:res_3}d). Despite the fact
that no explicit memory patterns are assigned to the the inhibitory
populations, the inhibitory activity plays a computational role that
goes beyond providing global inhibitory feedback; when the weights
of the I to E connections are shuffled, the network's performance
significantly degrades (Fig. S5). 

\subsection*{Learning Robust Solutions}

Thus far, we have presented analytical and numerical investigations
of solutions that support selectivity or associative memory and provide
substantial robustness to noise. However, we did not address the way
in which these robust solutions could be learned by a biological system.
In fact, as stated above, the majority of solutions for these tasks
have vanishingly small output and input robustness. Therefore, an
important question is whether noise robust weights can emerge naturally
from synaptic learning rules that are appropriate for neuronal circuits.

The actual algorithms used for learning in the neural circuits are
generally unknown, especially within a supervised learning scenario.
Experiments suggest that learning rules may depend on brain area and
both pre and post synaptic neuron types (see for example \cite{lu_spike-timing-dependent_2007,nugent_ltp_2008,chevaleyre_heterosynaptic_2003,damour_inhibitory_2015},
for reviews see \cite{mcbain_presynaptic_2009,kullmann_long-term_2007,lamsa_spike-timing_2010,larsen_synapse-type-specific_2015}).
Regardless of the particular learning algorithm used, however, our
theory suggests that a simple way to ensure that learning arrives
at a robust solution is to introduce noise during learning. Indeed,
this is a common practice in machine learning for increasing generalization
abilities (a specific form of \emph{data augmentation}, \cite{srivastava_dropout:_2014,lecun_deep_2015}).
The rationale is that learning algorithms that achieve low error in
the presence of noise necessarily lead to solutions that are robust
against noise levels at least as large as those present during learning.
In the case we are considering, learning in the presence of substantial
\emph{input noise} should lead to solutions that have substantial
$\kappa_{\mathrm{in}}$ and introducing \emph{output noise} during
learning should lead to solutions with substantial $\kappa_{\mathrm{out}}$.
We note that $\kappa_{\mathrm{in}}$ may be large even if $\kappa_{\mathrm{out}}$
remains small (for example, in unbalanced solutions with maximal $\kappa_{\mathrm{in}}$)
but not \emph{vice versa} (because $\kappa_{\mathrm{out}}$ of order
1 implies $\wnorm$ (and as a result $\kappa_{\mathrm{in}}$) of order
1 as well). Therefore, learning in the presence of significant output
noise should lead to solutions that are robust to both input and output
noise, whereas learning in the presence of input noise alone may lead
to unbalanced solutions that are sensitive to output noise, depending
on details of the learning algorithm. We therefore predict that performing
successful learning in the presence of output noise is a sufficient
condition for the emergence of excitation-inhibition balance. 

To demonstrate that robust balanced solutions emerge in the presence
of output noise, we consider a variant of the perceptron learning
algorithm \cite{rosenblatt_principles_1962} in which we have forced
the sign constraints on the weights \cite{amit_perceptron_1989} and,
in addition, added a weight decay term implementing a soft constraint
on the magnitude of the weights (\emph{Methods}). This supervised
learning rule possesses several important properties that are required
for biological plausibility: It is on-line, weights are modified incrementally
after each pattern presentation; It is history independent so that
each weight update only depends on the current pattern and error signal;
Lastly, it is simple and local, weight updates are a function of the
error signal and quantities that are available locally at the synapse
(presynaptic activity and synaptic efficacy). When this learning rule
is applied to train a selectivity task in the presence of substantial
output noise, the resulting solution has a balanced weight vector
with substantial $\kappa_{\mathrm{out}}$ and $\kappa_{\mathrm{in}}$
(Fig. \ref{fig:learn-1}, black). In contrast, if learning occurs
with weak output noise, the resulting solution is unbalanced with
small $\kappa_{\mathrm{out}}$, while its $\kappa_{\mathrm{in}}$
may be large if substantial input noise is present during learning
(Fig. \ref{fig:learn-1}, gray). When this learning rule is applied
in the load regime where only unbalanced solutions exist ($\alpha_{\mathrm{b}}<\alpha<\alpha_{\mathrm{c}}$),
learning fails to achieve reasonable performance when applied in the
presence of large output noise. When noise is scaled down to the value
allowed by $\kappa_{\mathrm{out}}^{\mathrm{max}}\propto1/\sqrt{N}$,
learning yields unbalanced solutions with robustness values of the
order of the maximum allowed in this region (Fig. S6). 

\begin{figure}
\begin{minipage}[t]{1\columnwidth}%
\begin{wrapfigure}{o}{0.525\columnwidth}%
\includegraphics{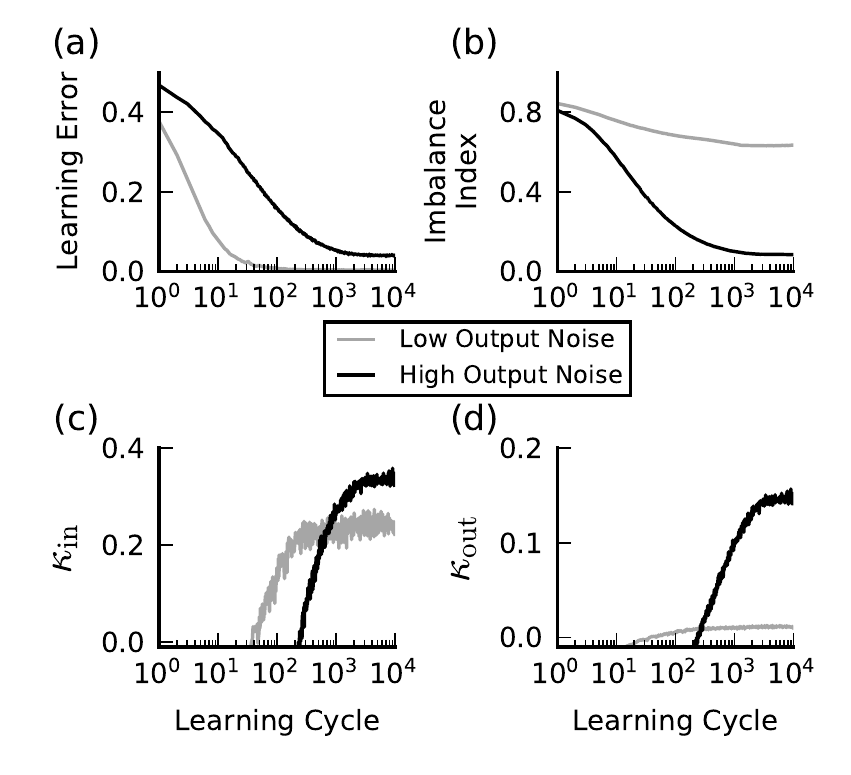}\end{wrapfigure}%
\noindent\refstepcounter{figure}Figure \thefigure: \label{fig:learn-1}\textbf{\small{}Emergence
of E-I balance from learning in the presence of output noise. }{\small{}All
panels show the outcome of perceptron learning for a noisy neuron
(}\emph{\small{}Methods}{\small{}) under low ($\sigma_{\mathrm{out}}=0.01$,
gray) and high output noise conditions ($\sigma\mathrm{_{out}}=0.1$,
black). Except for $\sigma_{\mathrm{out}}$, all model and learning
parameters are identical for the two conditions (including $\sigma_{\mathrm{in}}=0.1$).
(a) Mean training error vs. learning cycle. On each cycle, all the
input patterns to be learned are presented once. The error decays
and plateaus at its minimal value under both low and high output noise
conditions (b) Mean imbalance index (IB, eq. \ref{IB}) vs. learning
cycle. IB remains of order 1 under low output noise conditions and
drops close to zero under high output noise conditions. (c) Mean input
robustness ($\kappa_{\mathrm{in}}$) vs. learning cycle. Input robustness
is high under both output noise conditions. (d) Mean output robustness
($\kappa_{\mathrm{out}})$ vs. learning cycle. Output robustness is
substantial only under the high output noise learning condition. These
results demonstrate that robust balanced solutions naturally emerge
under learning in the presence of high output noise. See }\emph{\small{}Methods}{\small{}
for other parameters used.}%
\end{minipage}
\end{figure}

\section*{Discussion}

The results we have presented come from imposing a set of fundamental
biological constraints: fixed-sign synaptic weights, non-negative
afferent activities, a positive firing threshold (relative to the
resting potential), and both input and output forms of noise. Amit
et al.\ \cite{amit_interaction_1989} studied the maximal margin
solution for the sign-constrained perceptron and showed that it has
half the capacity of the unconstrained perceptron. However, this previous
work considered afferent activities that were centered around zero
and a neuron with zero firing threshold, features that preclude the
presence of the novel behavior exhibited by the more biologically
constrained model studied here. Chapeton et al. \cite{chapeton_efficient_2012}
studied perceptron learning with sign-constrained weights and a preassigned
level of robustness, but only considered solutions in the unbalanced
regime which, as we have shown, are extremely sensitive to output
noise. 

Learning in neural circuits involves a trade-off between exhausting
the system's capacity for implementing complex input-output functions
on the one hand, and ensuring good generalization properties on the
other. A well-known approach in machine learning has been to search
for solutions that fit the training examples while maximizing the
distance of samples from the decision surface, a strategy known as
maximizing the margin \cite{vapnik_nature_2000,gardner_space_1988,amit_interaction_1989}.
The margin being maximized in this case corresponds, in our framework,
to $\kappa_{\mathrm{in}}$. Work in computational neuroscience has
implicitly optimized a robustness parameter equivalent to our $\kappa_{\mathrm{out}}$
\cite{brunel_optimal_2004,chapeton_efficient_2012}. To our knowledge,
the two approaches have not been distinguished before nor shown to
result in solutions with dramatically different noise sensitivities.
In particular, over a wide parameter range, we have shown that maximizing
$\kappa_{\mathrm{out}}$ leads to a balanced solution with minimal
sensitivity to output noise and robustness to input noise that is
almost as good as that of the maximal margin solution, with only a
modest trade-off in capacity. On the other hand, maximizing the margin
($\kappa_{\mathrm{in}}$) often leads to unbalanced solutions with
extreme sensitivity to output noise. 

The perceptron has long been considered a model of cerebellar learning
and computation \cite{marr_theory_1969,albus_theory_1971}. More recently,
Brunel et al.\ \cite{brunel_optimal_2004} investigated the capacity
and robustness of a perceptron model of a cerebellar Purkinje cell,
taking all weights to be excitatory. In view of the analysis presented
here, balanced solutions are not possible in this case ($f_{\mathrm{exc}}=1$),
and solutions that maximize either input-noise or output-noise robustness
both have $\kappa_{\mathrm{out}}\propto1/\sqrt{N}$. These two types
of solutions differ in their weight distributions, with experimentally
testable consequences for the predicted circuit structure (\emph{SI
Methods} and Fig. S2c; Brunel et al.\  \cite{brunel_optimal_2004}
only considered solutions that maximize $\kappa_{\mathrm{out}}$).
Output robustness of the unbalanced solutions can be increased by
making the input activity patterns sparse. Denoting by $s$ the mean
fraction of active neurons in the input, maximum output robustness
scales as $\kappa_{\mathrm{out}}\sim1/\sqrt{Ns}$ (Fig. \ref{fig:res_1}b,
and \emph{SI Methods}).\textcolor{black}{{} Thus, the high sparsity
in input activation (granule cell activity) of the cerebellum relative
to the modest sparsity in the neocortex is consistent with the former
being dominated by excitatory modifiable synapses. }

Interestingly, our results suggests an optimal ratio of excitatory
to inhibitory synapses. Capacity in the balanced regime is optimal
when $f_{\mathrm{exc}}=f_{\mathrm{exc}}^{\star}$, with $f_{\mathrm{exc}}^{\star}$
determined by the coefficients of variation (with respect to stimulus)
of the excitatory and inhibitory inputs (eq. \ref{f_star}). Thus,
optimality predicts a simple relation between the fraction of excitatory
and inhibitory inputs and their degree of tuning. Estimating the CV's
from existing data is difficult, but it would be interesting to check
if input statistics and connectivity ratios in different brain areas
are consistent with this prediction. The commonly observed value in
cortex, $f_{\mathrm{exc}}\simeq0.8$, would be optimal for input statistics
with $\mathrm{CV_{exc}}/\mathrm{CV_{inh}}$$\simeq4$. In general,
we expect that $\mathrm{CV_{exc}}/\mathrm{CV_{inh}}>1$ which implies
that $f_{\mathrm{exc}}^{\star}>1/2$.

For most of our work, we assumed that inhibitory neurons learn to
represent specific sensory and long-term memory information, just
as the excitatory ones and that all synaptic pathways are learned
using similar learning rules. While plasticity in both excitatory
and inhibitory pathways have been observed \cite{lu_spike-timing-dependent_2007,nugent_ltp_2008,chevaleyre_heterosynaptic_2003,damour_inhibitory_2015,lamsa_spike-timing_2010,larsen_synapse-type-specific_2015,hennequin_inhibitory_2017},
accumulating experimental evidence indicates a high degree of cell
type and synaptic type specificity of the plasticity rules. In addition,
synaptic plasticity is under tight control of neuromodulatory systems.
At present, it is unclear how to interpret our learning rules in terms
of concrete experimentally observed synaptic plasticity. Other functional
models of neural learning assume learning only within excitatory population
with inhibition acting as global stabilizing force. In the case of
sensory processing, our approach is consistent with the observation
of a similar stimulus tuning of EPSCs and IPSCs in many cortical sensory
areas. The role of inhibitory neurons in memory representations is
less known (but see \cite{wilent_discrete_2007}). Importantly, we
have shown that our main results are valid also in the case in which
inhibitory neurons do not explicitly participate in the coding of
the memories. Interestingly, our work suggests that even if inhibitory
neurons are only passive observers during learning processes, learning
of inhibitory synapses onto excitatory cells can amplify the memory
stability of the system against fluctuations in the inhibitory feedback.
Given the diversity of inhibitory cell types it is likely that in
the real circuits inhibition plays multiple roles, including both
conveying information and providing stability. 

Several previous models of associative memory have incorporated biological
constraints on the sign of the synapses, Dale's Law, assuming variants
of Hebbian plasticity in the E to E synapses \cite{amit_associative_1989,golomb_willshaw_1990,barkai_modulation_1994,hasselmo_acetylcholine_1993,van_vreeswijk_course_2005,roudi_balanced_2007}.
The capacity of the these Hebbian models is relatively poor, and their
basins of attractions small, except at extremely sparse activity levels.
In contrast, our model applies a more powerful learning rule that,
while keeping the sign constraints on the synapses, exhibits significantly
superior performance: with high capacity even for moderate sparsity
levels, large basins of attraction and high robustness to output noise. 

From a dynamical systems perspective, the associative memory networks
we construct exhibit unusual properties. In most associative memory
network models large basins of attractions endow the memory state
with robustness against stochasticity in the dynamics (\emph{i.e.},
output noise). Here, we found that, for the same set of fixed-point
memories, the synaptic weights with the largest possible basins (the
unbalanced solutions with maximal $\kappa_{\mathrm{in}}$) are very
sensitive to even mild levels of stochasticity, whereas the balanced
synaptic weights with somewhat reduced basins have substantially increased
output noise robustness.

At the network level, as at the single-neuron level, imposing basic
features of neural circuitry \textendash{} positive inputs, bounded
synapses of fixed sign, a positive firing threshold, and sources of
noise \textendash{} force neural circuits into the balanced regime.
A recent class of models showing computational benefits of balanced
inputs use extremely strong synapses, which are outside the range
we have discussed \cite{boerlin_predictive_2013}. These models are
stabilized by instantaneous transmission of signals between neurons
which are not required in the range of synaptic strength we consider. 

Previous models of balanced networks have highlighted the ability
of network with strong excitatory and inhibitory recurrent synapses
to settle into a state in which the total input is dynamically balanced
without special tuning of the synaptic strengths. Such a state is
characterized by a high degree of intrinsically generated spatio-temporal
variability \cite{van_vreeswijk_chaos_1996}. Mean population activities
respond fast and in a linear fashion to external inputs. Typically,
these networks lack the population level nonlinearity required to
generate multiple attractors. In contrast, we have explored the capacity
of balanced network to support multiple stable fixed-points by tuning
the synaptic strengths through appropriate learning. Despite the dynamic
and functional differences in the two classes of networks, the balancing
of excitation and inhibition plays a similar role in both. In the
first scenario, synaptic balance amplifies small changes in the spatial
or temporal properties of the external drive. Similarly, in the present
scenario, balanced synaptic architecture leads to enhanced robustness
by amplifying the small variations in the synaptic inputs induced
by changes in the stimulus or memory identity. It would be interesting
to combine fast dynamics with robust associative memory capabilities. 

In conclusion, we have uncovered a fundamental principle of neuronal
learning under basic biological constraints. Our work reveals that
excitation-inhibition balance may have a critical computational role
in producing robust neuronal functionality that is insensitive to
output noise. We showed that this balance is important at the single
neuron level for both spiking and non-spiking neurons, and at the
level of recurrently connected neural networks. Further, the theory
suggests that excitation-inhibition balance may be a collective, self-maintaining,
emergent phenomena of synaptic plasticity. Any successful neuronal
learning process in the presence of substantial output noise will
lead to strong balanced synaptic efficacies with noise robustness
features. The fundamental nature of this result suggests that it should
apply across a variety of neuronal circuits that learn in the presence
of noise.

\section*{Methods}

\subsection*{Finding perceptron solutions}

There are a number of numerical methods for choosing a weight vector
$\w$ that generates a specified selectivity \cite{amit_perceptron_1989,brunel_optimal_2004,chapeton_efficient_2012,vapnik_nature_2000}.
For numerical simulations we developed algorithms that find the maximal
$\kappa_{\mathrm{out}}$ and maximal $\kappa_{\mathrm{in}}$ solutions
that obey the imposed biological constraints. These solutions can
be found directly by solving conic programing optimization problems
for which efficient algorithms exist and are widely available \cite{andersen2011}.
For details see \emph{SI Methods}.

\subsection*{Simulations of recurrent networks}

\textbf{Memory states:} Networks were trained to implement a set of
$P$ memory states, specified by $x_{i}^{\mu}\in\left\{ 0,1\right\} ,$
$i=1,2,\dots,N$, $\mu=1,2,\dots,P$, as stable fixed points of the
noise free dynamics. Memory states were randomly chosen i.i.d. from
binary distributions with parameter $p_{\mathrm{exc/inh}}$ according
to the type of the $i$'th input afferent, \emph{i.e.}, $\mathrm{Pr}\left(x_{i}^{\mu}=1\right)=p_{\mathrm{exc}/\mathrm{inh}}$
and $\mathrm{Pr}\left(x_{i}^{\mu}=0\right)=1-p_{\mathrm{exc}/\mathrm{inh}}$.
\textbf{Initial pattern distortion: }To start the network close to
a memory state $\boldsymbol{x}^{\mu}$, the initial state of the network,
$s_{i}\left(t=0\right)$ for $i=1,2,\dots,N$, was randomly chosen
according to $\mathrm{Pr}\left(s_{i}=1\right)=\left(1-\delta\right)\Theta\left(2x_{i}^{\mu}-1\right)+\delta\frac{p_{\mathrm{exc/inh}}}{1-p_{\mathrm{exc/inh}}}\Theta\left(-2x_{i}^{\mu}+1\right)$
where $\delta$ is the initial pattern distortion level (Fig. \ref{fig:res_3}b)
and $\Theta\left(x\right)=1$ for $x\geq0$ and 0 otherwise. This
procedure ensures that the mean activity levels of excitatory and
inhibitory neurons in the initial state is the same as the their mean
activity levels in the memory state \cite{litwin-kumar_optimal_2017}.

\subsection*{Perceptron learning algorithm}

The perceptron learning algorithm learns to classify a set of $P$
labeled patterns. At learning time step $t$ one pattern $\boldsymbol{x}_{t}$
with desired output $y_{t}=\pm1$ is presented to the neuron. The
output of the perceptron $s_{t}$ is given by $s_{t}=\mathrm{sign}\left(\boldsymbol{w}_{t}^{\mathrm{T}}\boldsymbol{x}_{t}+\eta_{t}-1\right)\ ,$
where $\eta_{t}$ is a Gaussian random variable with zero mean and
variance $\left|\boldsymbol{w}_{t}\right|^{2}\sigma_{\mathrm{in}}^{2}+\sigma_{\mathrm{out}}^{2}$.
The error signal is defined as $e_{t}=y_{t}\Theta\left(-s_{t}y_{t}\right)$
where $\Theta\left(x\right)=1$ for $x>0$ and zero otherwise. After
each pattern presentation all synapses are updated. The synaptic weights
of excitatory inputs are updated according to $w_{i,t+1}=\left[\left(1-\varepsilon\right)w_{i,t}+\rho e_{t}x_{i,t}\right]_{+}$
and weights of inhibitory inputs are updated according to $w_{i,t+1}=\left[\left(1-\varepsilon\right)w_{i,t}+\rho e_{t}x_{i,t}\right]_{-}$
where $\left[x\right]_{\pm}=x\Theta\left(\pm x\right)$, $\varepsilon$
is a weight decay constant and $\rho$ is a constant learning rate.
At each learning cycle ($P$ learning time steps) all patterns are
presented sequentially in a random order (randomized at each learning
cycle). 

\subsection*{Random patterns in numerical estimation of $\kappa_{\mathrm{out}}^{\max}$
and $\kappa_{\mathrm{in}}^{\mathrm{max}}$ solutions }

In numerical experiments for Fig. 3, Fig. 4, Fig. S1 and Fig. S3,
excitatory inputs for the random patterns were drawn i.i.d from an
exponential distribution with unity mean and standard deviation. Inhibitory
inputs were drawn from a Gamma distribution with shape parameter $k$
and scale parameter $\theta$ (The PDF of the Gamma distribution is
given by $P\left(x\right)=\frac{1}{\Gamma\left(k\right)\theta}\left(\frac{x}{\theta}\right)^{k-1}e^{-\frac{x}{\theta}}$
where $\Gamma\left(k\right)$ is the Gamma function). 

\subsection*{Dynamics of Leaky Integrate-and-Fire neuron}

\textbf{Input spike-trains:} For each input pattern $\boldsymbol{x}^{\mu}$
input spike trains of input afferent $i=1,2,...,N$ were drawn randomly
from a Poisson processes with rate $r_{i}=Ax_{i}^{\mu},$ for duration
$T$. \textbf{Synaptic Input: }Given the set of input spike trains
$\left\{ t_{i}\right\} ,$ $i=1,2,...,N$ the contribution of synaptic
input to the membrane potential is given by $V_{\mathrm{syn}}\left(t\right)=\sum_{i}w_{i}\sum_{t_{i}}K\left(t-t_{i}\right)$
where $w_{i}$ is the synaptic efficacy of the synapse from the $i$'th
input afferent and $K\left(t\right)$ is a post synaptic potential
kernel. $K\left(t\right)=0$ for $t<0$ and is given by $K\left(t\right)=V_{0}\left(e^{-\frac{t}{\tau_{\mathrm{m}}}}-e^{-\frac{t}{\tau_{\mathrm{s}}}}\right)$
for $t>0$ where $\tau_{\mathrm{m}}$ and $\tau_{\mathrm{s}}$ are
the membrane and synaptic time constants respectively, and $V_{0}$
is such that the maximal value of $K\left(t\right)$ is one \textbf{Output
noise:} Output noise was added to the neuron's membrane potential
as random synaptic input $V_{\mathrm{o.n.}}\left(t\right)=\sum_{j=1}^{N_{\mathrm{noise}}}g_{j}K\left(t-t_{j}\right)$
were $g_{j}$ was randomly drawn from a zero mean Gaussian distribution
with standard deviation $\sigma_{\mathrm{n}}$ and $t_{j}\in\left(0,T\right)$
was randomly drawn from a uniform distribution.\textbf{ Voltage reset:
}After each threshold crossing the membrane potential was reset to
it's resting potential. Given the set of output spike times $\left\{ t_{\mathrm{spike}}\right\} $,
the total contribution of voltage reset to the membrane potential
can be written as $V_{\mathrm{reset}}\left(t\right)=-\left(V_{\mathrm{th}}-V_{\mathrm{rest}}\right)\sum_{t_{\mathrm{spike}}}R\left(t-t_{\mathrm{spike}}\right)$
where $V_{\mathrm{rest}}$ and $V_{\mathrm{th}}$ are the neuron's
resting and threshold potential respectively and $R\left(t\right)$
implements the post-spike voltage reset. $R\left(t\right)=0$ for
$t<0$ and is given by $R\left(t\right)=e^{-\frac{t}{\tau_{\mathrm{m}}}}$
for $t\ge0$. This form ensures the voltage is reset to the resting
potential immediately after an output spike. \textbf{Membrane potential:
}Finally, the neuron's membrane potential is given by $V\left(t\right)=V_{\mathrm{rest}}+V_{\mathrm{syn}}\left(t\right)+V_{\mathrm{o.n.}}\left(t\right)+V_{\mathrm{reset}}\left(t\right)$
where $V_{\mathrm{reset}}$ is computed given $V_{\mathrm{syn}}$
and $V_{\mathrm{o.n.}}$. 

\subsection*{Figures parameters}

\textbf{Fig. \ref{fig:res_2}: }In all panels $\sigma_{\mathrm{inh}}/\sigma_{\mathrm{exc}}=2$
and $\mathrm{CV_{exc}}/\mathrm{CV_{inh}}=\sqrt{2}$. $\sigma_{\mathrm{inh}}/\sigma_{\mathrm{exc}}=2,\ \mathrm{CV_{exc}}/\mathrm{CV_{inh}}=\sqrt{2}$
with an even split between responsive/unresponsive labels. In (b)
and (c) $f_{\mathrm{exc}}=0.8$. \textbf{Fig. \ref{fig:res_1}: }In
all panels $N=3000$, $k=2$ and $\theta=\sqrt{2}$ leading to $\sigma_{\mathrm{inh}}/\sigma_{\mathrm{exc}}=2\ \mathrm{CV_{exc}}/\mathrm{CV_{inh}}=\sqrt{2}$
, $f_{\mathrm{exc}}=0.8$ with an even split between responsive/unresponsive
labels. Numerical results are averaged over 100 samples. \textbf{Fig.
\ref{fig:spikes}:} In all panels $N=1000$, $P=1000$, fraction of
`plus' patterns $p_{\mathrm{out}}=0.1,$ $f_{\mathrm{exc}}=0.8$,
$V_{\mathrm{rest}}=0$, $V_{\mathrm{thr}}=1$, $\tau_{\mathrm{m}}=30\mathrm{msec},$
$\tau_{\mathrm{s}}=10\mathrm{msec}$, $T=200\mathrm{msec},$ $A=30\mathrm{Hz}.$
Random patterns were drawn as described above with $k=2$ and $\theta=\sqrt{2}$.
Maximal $\kappa_{\mathrm{out}}$ solutions were found with $\Gamma=1.5$
in units of $\left(V_{\mathrm{th}}-V_{\mathrm{rest}}\right)/\sigma_{\mathrm{exc}}$.
No output noise was added in panels (a)-(c). In panels (d)-(e) output
noise was added with $N_{\mathrm{noise}}=30,000$ and\textbf{ $\sigma_{\mathrm{n}}=2/\sqrt{N_{\mathrm{noise}}}$}
(see above).\textbf{ Fig. \ref{fig:res_3}: }In panels (c)-(f) $N=2000$,
$P=1000$, $f_{\mathrm{exc}}=0.8$, $p_{\mathrm{exc}}=0.1$, $p_{\mathrm{inh}}=0.2$,
$\Gamma=10\sqrt{p_{\mathrm{exc}}\left(1-p_{\mathrm{exc}}\right)}$
in units of $\left(V_{\mathrm{th}}-V_{\mathrm{rest}}\right)/\sigma_{\mathrm{exc}}$.
In (d) and (f) results are averaged over 10 networks and 10 patterns
from each network. See \emph{SI Methods }for parameters of inhibitory
connectivity of the non learned inhibition networks (gray lines).
In panel (g) maximal output noise magnitude is defined as the value
of $\sigma_{\mathrm{out}}$ for which the stable pattern probability
is 1/2. To minimize finite size effects in simulations we used $N=3000$,
$f_{\mathrm{exc}}=0.8,$ $p_{\mathrm{exc}}=0.5$, $p_{\mathrm{inh}}=0.5$,
$\Gamma=10\sqrt{p_{\mathrm{exc}}\left(1-p_{\mathrm{exc}}\right)}$
in units of $\left(V_{\mathrm{th}}-V_{\mathrm{rest}}\right)/\sigma_{\mathrm{exc}}$.
Stable pattern probability for each load and noise level was estimated
by averaging over 5 networks and 20 patterns from each network. \textbf{Fig.
\ref{fig:learn-1}: }Random patterns are binary pattern $x_{i}^{\mu}\in\left\{ 0,1\right\} $
with equal probabilities and an even split of `plus' and `minus' patterns.
$N=3000$, $P=900$. Learning algorithm parameters: $\varepsilon=5\cdot10^{-7}$,
$\rho=0.1/N$, $\sigma_{\mathrm{in}}=0.1$. Results are averaged over
50 samples.

\subsection*{Software}

To acknowledge their contribution to scientific work we cite the open
source projects that directly and most crucially contributed to the
current work. All computational aspects of this work were done using
the Python stack of scientific computing (CPython, Numpy, Scipy, Matplotlib
\cite{Hunter:2007}, Jupyter/Ipython \cite{PER-GRA:2007} and others).
Convex conic optimization was performed using CVXOPT \cite{andersen2011}.
Parallelization of simulations on a cluster computer was performed
using IPyparallel.

\subsection*{Code availability}

Python code for simulations and numerical solution of saddle point
equations is available upon request.

\section*{Acknowledgments }

We thank Misha Tsodyks for helpful discussions. Research supported
by NIH grant MH093338 (L.F.A. and R.R.), the Gatsby Charitable Foundation
through the Gatsby Initiative in Brain Circuitry at Columbia University
(L.F.A. and R.R.) and the Gatsby Program in Theoretical Neuroscience
at the Hebrew University (H.S), the Simons Foundation (L.F.A., R.R.
and H.S), the Swartz Foundation (L.F.A., R.R. and H.S), and the Kavli
Institute for Brain Science at Columbia University (L.F.A. and R.R.).

\bibliographystyle{pnas2011}
\bibliography{Rubin2017Balanced,SCP_comp}

\global\long\def\xbar{\bar{x}}
\global\long\def\ex{\mathrm{exc}}
\global\long\def\inh{\mathrm{inh}}
\global\long\def\obar{\bar{\omega}}
\global\long\def\obarhat{\hat{\bar{\omega}}}
\global\long\def\qhat{\hat{q}}
\global\long\def\wnorm{\left|\boldsymbol{w}\right|}
\global\long\def\Uthr{U_{\mathrm{thr}}}
\global\long\def\order#1{\mathcal{O}\left(#1\right)}
\global\long\def\kin{\kappa_{\mathrm{in}}}
\global\long\def\kout{\kappa_{\mathrm{out}}}
\global\long\def\kinmax{\kappa_{\mathrm{in}}^{\mathrm{max}}}
\global\long\def\koutmax{\kappa_{\mathrm{out}}^{\mathrm{max}}}

\part*{Supplementary Information: Balanced Excitation and Inhibition are
Required for High-Capacity, Noise-Robust Neuronal Selectivity}

Ran Rubin, L.F. Abbott and Haim Sompolinsky

\newpage{}

\tableofcontents{}

\selectlanguage{american}%
\renewcommand{\thefigure}{S\arabic{figure}}
\renewcommand{\theequation}{S.\arabic{equation}} 
\setcounter{figure}{0}
\setcounter{equation}{0}

\pagebreak{}
\selectlanguage{english}%

\section{Supplementary Figures}

\selectlanguage{american}%
\pagebreak{}\foreignlanguage{english}{}
\begin{figure}[H]
\selectlanguage{english}%
\includegraphics[width=1\textwidth]{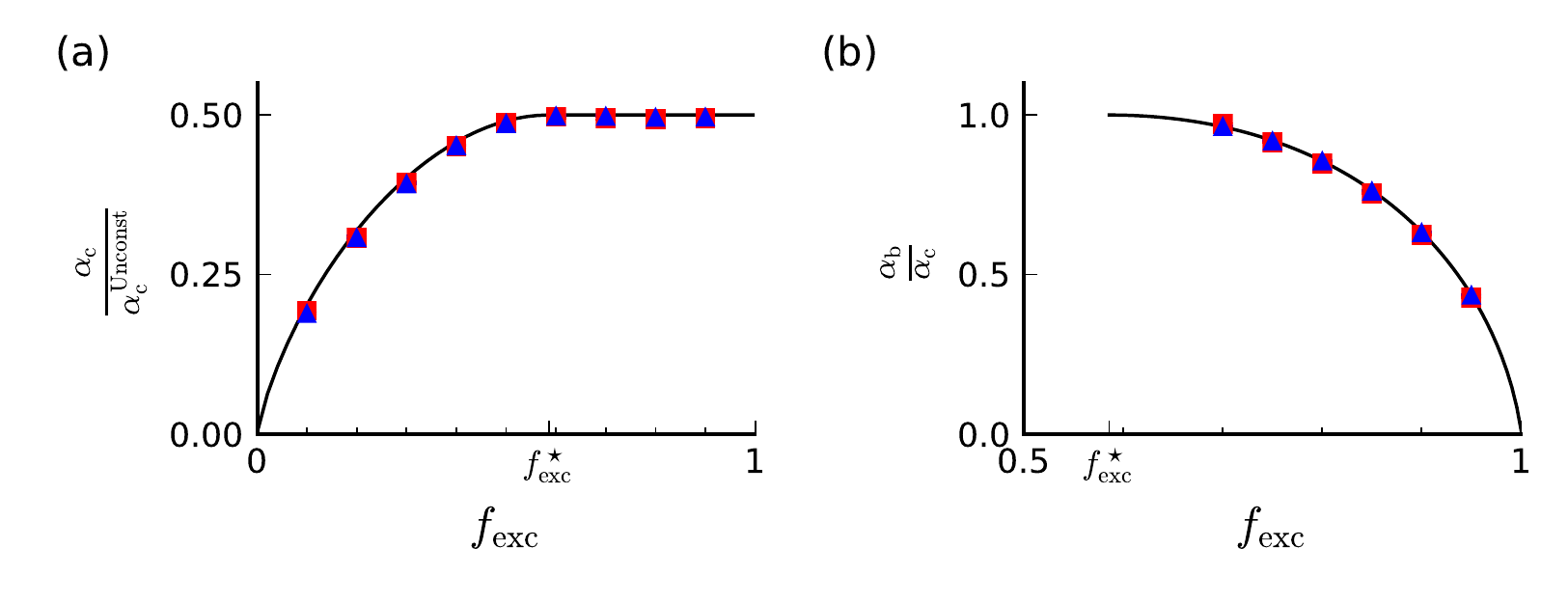}

\caption{\label{fig:res_1-1}\textbf{Numerical measurement of capacity and
balanced capacity. }(a) Capacity of sign constrained weights Perceptron,
$\alpha_{\mathrm{c}}$, vs. the fraction of excitatory inputs, $f_{\protect\ex}$,
as a fraction of the capacity of an unconstrained Perceptron (see
section \ref{sec:Capacity-for-non-even}). Theory is depicted in black.
Simulations results are shown in blue for $p_{\mathrm{out}}=0.5$
and red for $p_{\mathrm{out}}=0.1$. To measure $\alpha_{\mathrm{c}}$
we measure the probability of the existence of a solution as a function
of $\alpha$. We estimate $\alpha_{\mathrm{c}}$ by the load at which
this probability is 1/2. (b) Capacity of balanced solutions, $\alpha_{\mathrm{b}}$,
as a fraction of $\alpha_{\mathrm{c}}$ vs. $f_{\protect\ex}>f_{\mathrm{exc}}^{\star}$.
Since $\protect\koutmax$ solutions are balanced whenever balanced
solutions exist, to measure $\alpha_{\mathrm{b}}$ we measure the
probability of finding a balanced $\protect\koutmax$ solution i.e.
a solution that saturates the upper bound on $\protect\wnorm$. We
estimate $\alpha_{\mathrm{b}}$ by the load at which this probability
is 1/2. In both panels, $N=3000$, $\mathrm{CV}_{\protect\ex}/\mathrm{CV}_{\protect\inh}=\sqrt{2}$. }
\selectlanguage{english}%
\end{figure}

\pagebreak{}

\selectlanguage{english}%
\begin{figure}[H]
\includegraphics[width=1\textwidth]{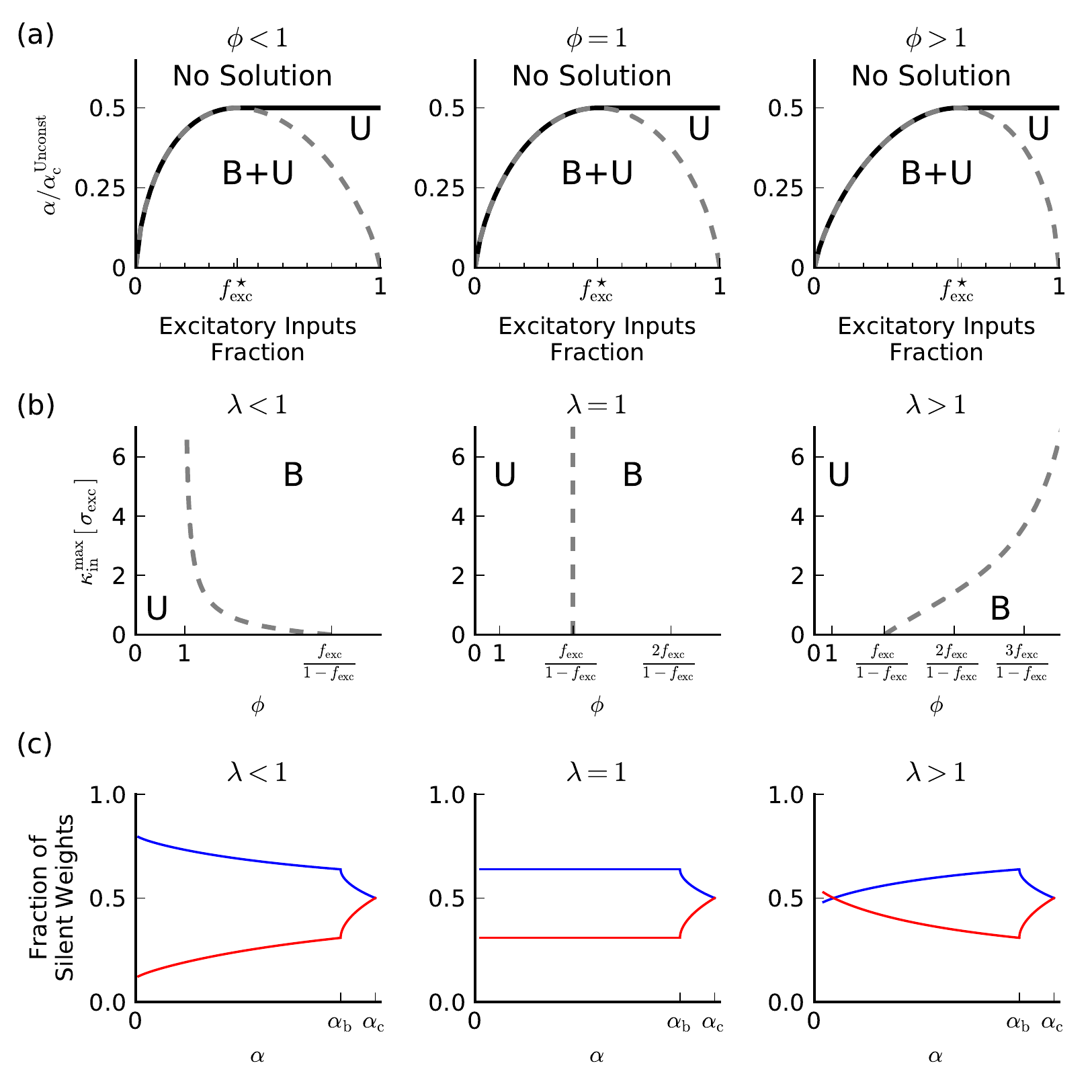}

\caption{\label{fig:input_stats}\textbf{Effects of input statistics.} (a)
Solution type vs. $f_{\protect\ex}$ and $\alpha$ (as a fraction
of $\alpha_{\mathrm{c}}^{\mathrm{Unconst}}$, see section \ref{sec:Capacity-for-non-even})
for different values of $\phi=\mathrm{CV_{\mathrm{exc}}/CV_{inh}}$.
From left to right $\phi=1/\sqrt{2},\ 1,\ \sqrt{2}$. Lines as in
Fig. 2a. (b) Type of \textbf{maximal $\protect\kin$ solutions} vs.
$\phi$ and $\protect\kinmax$ for different values of $\lambda=\sigma_{\mathrm{inh}}/\sigma_{\mathrm{\protect\ex}}$.
For a wide range of $\phi$ and $\lambda$ these solutions are unbalanced
for all values of $\protect\kinmax$. Here $f_{\protect\ex}=0.8$
and $\lambda=1/2,\ 1,\ 2$ from left to right. (c) Fraction of silent
weights for \textbf{maximal $\protect\kout$ solutions} vs. the load
for different values of $\lambda$. Fraction of silent excitatory
weights is shown in blue and fraction of inhibitory silent weights
is depicted in red. Here $f_{\protect\ex}=0.8,\ \phi=\sqrt{2}$, $p_{\mathrm{out}}=0.1$
and $\lambda=1/2,\ 1,\ 2$ from left to right. Notably, for unbalanced,
maximal $\protect\kin$ solutions the fraction of silent weights is
constant and equals 0.5 for both excitatory and inhibitory inputs
(not shown in figure).}
\end{figure}

\selectlanguage{american}%
\pagebreak{}

\selectlanguage{english}%
\begin{figure}[H]
\includegraphics[width=1\textwidth]{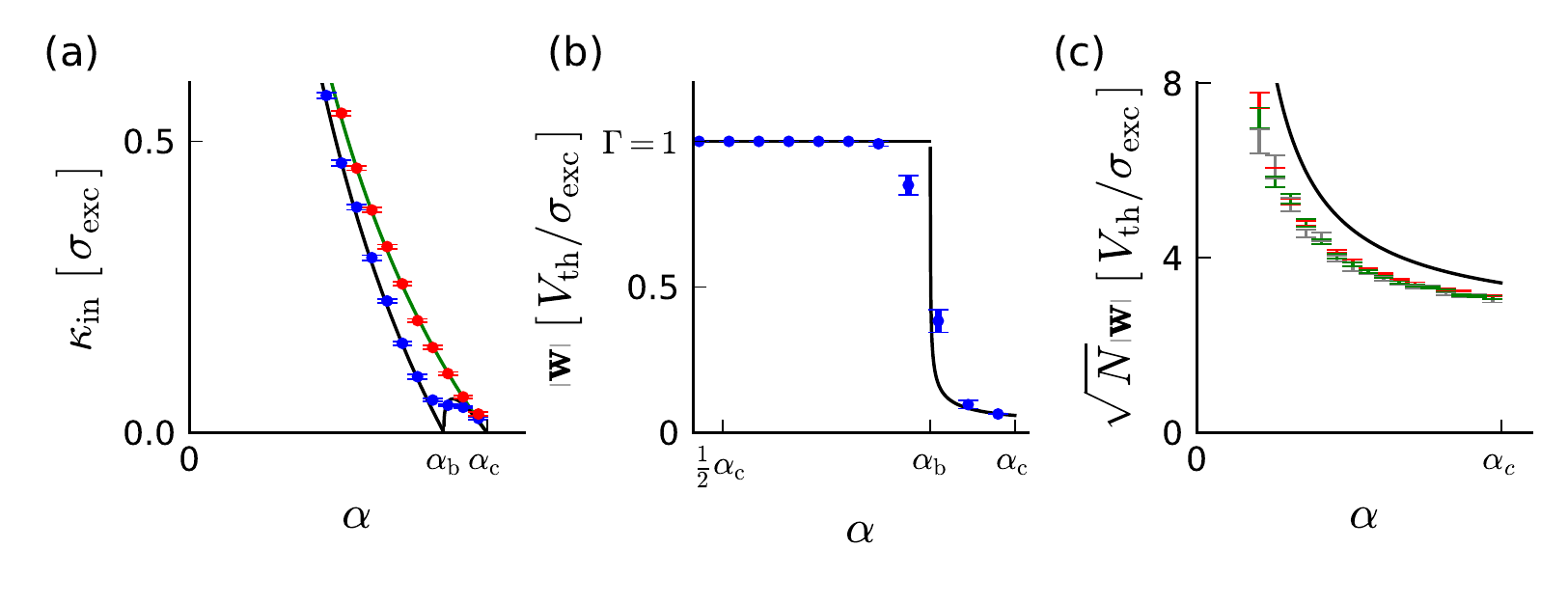}

\caption{\textbf{Properties of maximal output and input robustness solutions.
}(a) Input robustness, $\protect\kin$, vs the load for the maximal
$\protect\kin$ solution (red) and the maximal $\protect\kout$ solution
(blue). (b) Norm of synaptic weight vector vs. the load for the maximal
$\protect\kout$ solution. In the balanced regime ($\alpha<\alpha_{\mathrm{b}}$)
the norm saturates its upper bound $\Gamma=1$. Since the norm is
constant, maximizing $\protect\kout$ in the balanced regime is equivalent
of maximizing $\protect\kin$ under the constraint $\protect\wnorm=\Gamma$
(c) Rescaled norm of synaptic weight vector ($\sqrt{N}\protect\wnorm$)
vs. the load for the maximal $\protect\kin$ solution. To demonstrate
the $1/\sqrt{N}$ scaling of the weight vector norm, colors depict
results for $N=750$ (gray), $N=1500$ (green) and $N=3000$ (red).
In all panels lines depict theoretical prediction. $f_{\mathrm{exc}}=0.8$,
$p_{\mathrm{out}}=0.1$, $\phi=\mathrm{CV_{\mathrm{exc}}/CV_{inh}}=\sqrt{2},$
$\lambda=\sigma_{\mathrm{inh}}/\sigma_{\mathrm{\protect\ex}}=2$,
results are averaged over 100 samples. }
\end{figure}

\selectlanguage{american}%
\pagebreak{}

\selectlanguage{english}%
\begin{figure}
\includegraphics[width=1\columnwidth]{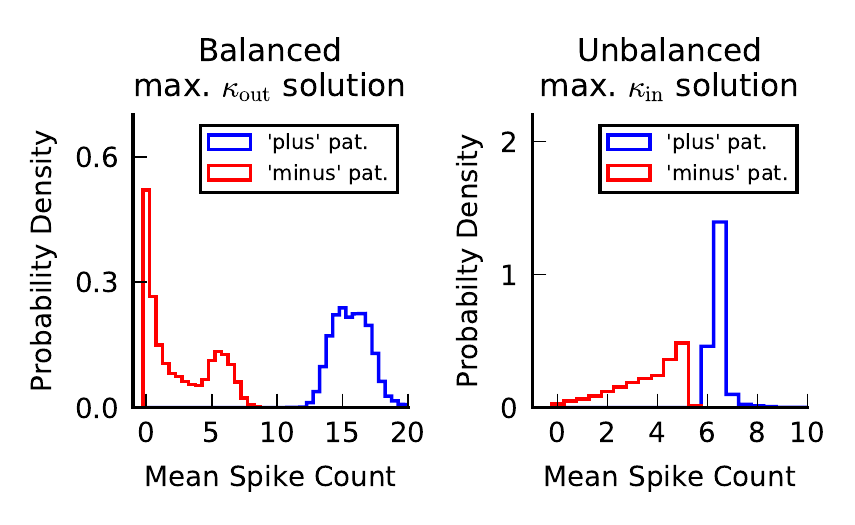}

\caption{\textbf{Neuronal selectivity for a spiking neuron. }Both panels depict
the histograms of the mean output spike count for patterns belonging
to the `plus' (blue) and `minus' (red) classes of an LIF neuron with
balanced weights maximizing $\kappa_{\mathrm{out}}$ (left) and unbalanced
weights maximizing $\kappa_{\mathrm{in}}$. Here the magnitude of
the output noise is zero. In both cases the mean output spike count
can be used to correctly classify the patterns. For parameters used
see Fig. 3.}

\end{figure}

\selectlanguage{american}%
\pagebreak{}

\selectlanguage{english}%
\begin{figure}
\includegraphics[width=1\textwidth]{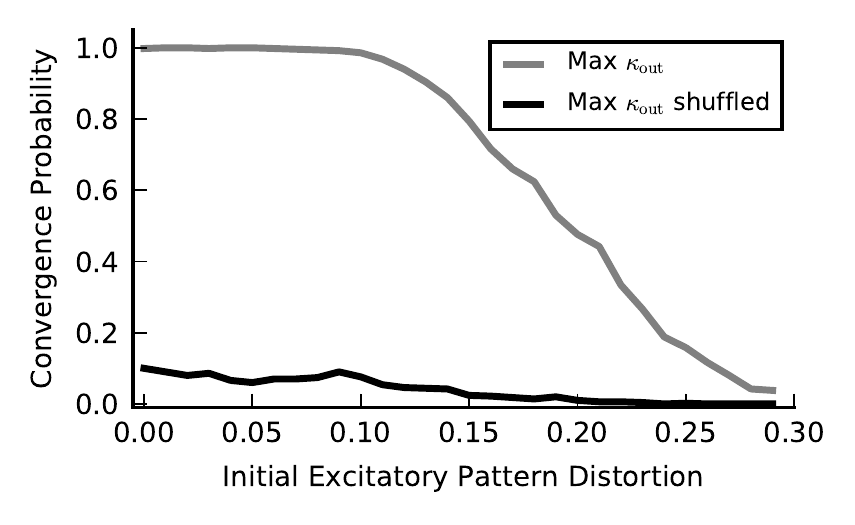}

\caption{\textbf{\label{fig:Effect-of-shuffling}Effect of shuffling learned
inhibitory weights in recurrent networks with non learned inhibitory
activity.} Gray line depicts the performance of the a network with
random E to I and I to I connection and learned E to E and I to E
connections (see section \ref{sec:non_learned_inh}, same as gray
line in Fig. 5d). Black line depicts the performance of the same network
with the inhibitory weights of each excitatory neuron randomly shuffled.
Thus the distribution of inhibitory synaptic weights for each excitatory
neurons is identical in both cases. This results shows that the learned
inhibitory weight are important for network performance and stability.}

\end{figure}

\selectlanguage{american}%
\pagebreak{}

\selectlanguage{english}%
\begin{figure}[H]
\includegraphics[width=0.75\textwidth]{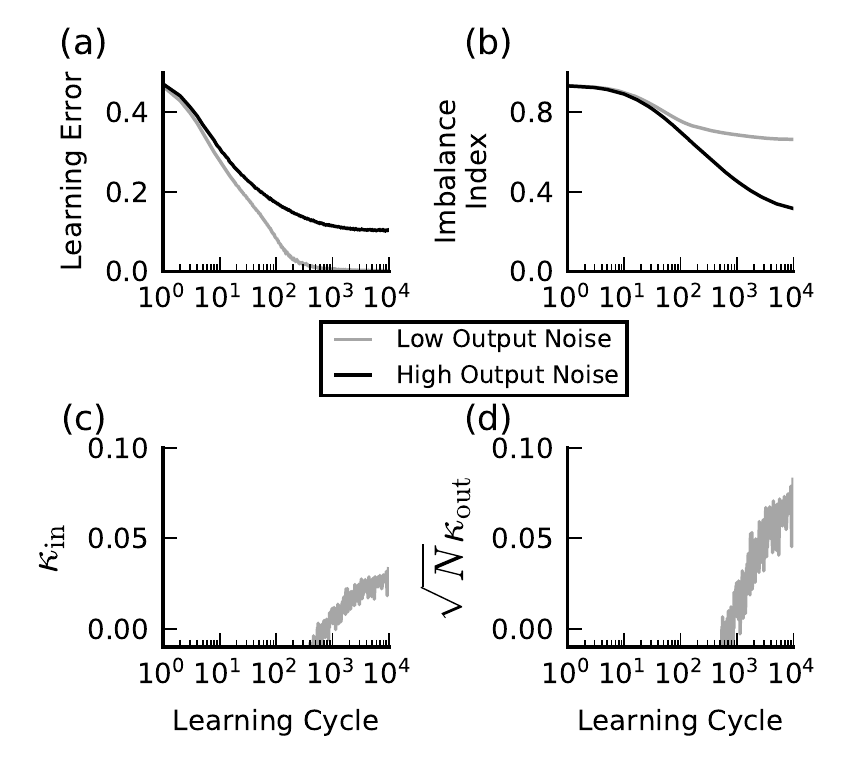}

\caption{\label{fig:learn-1}\textbf{Perceptron learning with input and output
noise for $\alpha_{\mathrm{b}}<\alpha<\alpha_{\mathrm{c}}$.} All
panels depict the outcome of simple perceptron learning for a noisy
neuron (Methods) under low output noise conditions ($\sigma_{\mathrm{out}}=0.01/\sqrt{N}$,
black) and high output noise conditions ($\sigma\mathrm{_{out}}=0.01$,
gray). Except $\sigma_{\mathrm{out}}$ all model and learning parameters
are identical for the two conditions (including $\sigma_{\mathrm{in}}=0.01$).
(a) Mean training error vs. learning cycle. At each cycle all the
learned input patterns are presented once. (b) Mean imbalance index
vs. learning cycle. IB remains of order 1 under low output noise conditions
and drops to lower values under high output noise conditions. (c)
Mean input robustness ($\kappa_{\mathrm{in}}$) vs. learning cycle.
(d) Mean rescaled output robustness ($\sqrt{N}\kappa_{\mathrm{out}}$)
vs. learning cycle. The error decays and plateaus at its minimal value
under both low and high output noise conditions, however for high
output noise the error remains substantial. Both output and input
robustness are negative under the high output noise conditions (The
learning does not find a weights vector that performs the classification
of the noise free patterns correctly). Input and output robustness
are positive when the output noise scales at most as $1/\sqrt{N}$.
Random patterns are binary pattern $x_{i}^{\mu}\in\left\{ 0,1\right\} $
with equal probabilities and an even split of `plus' and `minus' patterns.
$N=3000$, $P=2400$. Learning algorithm parameters: $\varepsilon=10^{-8}$,
$\rho=0.02/N$, $\sigma_{\mathrm{in}}=0.01$. Results are averaged
over 50 samples.}
 
\end{figure}

\selectlanguage{american}%
\pagebreak{}
\selectlanguage{english}%

\section{Finding maximal $\kappa_{\mathrm{in}}$ and maximal $\kappa_{\mathrm{out}}$
solutions}

Here we describe how finding the maximal $\kappa_{\mathrm{in}}$ and
maximal $\kappa_{\mathrm{out}}$ solutions can be expressed as convex
conic optimization problems. This allows us to efficiently validate
the theoretical results. As noted in the main text, maximizing $\kin$
is equivalent to maximizing the margin of the solution's weight vector
as is done by Support Vector Machines \cite{vapnik_nature_2000}.
However, to our knowledge, the application of conic optimization tools
for maximizing $\kout$ is a novel contribution of our work. 

Solution weight vectors, $\boldsymbol{w}$, with input robustness
$\kappa_{\mathrm{in}}$, or output robustness $\kappa_{\mathrm{out}}$,
satisfy the following inequalities: 
\begin{equation}
\forall\mu\ y^{\mu}\left(\boldsymbol{w}^{T}\boldsymbol{x}^{\mu}-V_{\mathrm{th}}\right)\ge D
\end{equation}
where $D_{\mathrm{in}}=\wnorm\kappa_{\mathrm{in}}$ and $D_{\mathrm{out}}=\kappa_{\mathrm{out}}$
(here we assume without loss of generality that $V_{\mathrm{rest}}=0$). 

For each solution $\boldsymbol{w}$ we define effective weights, $\boldsymbol{u}$,
and effective threshold $b$ (the so called canonical weights and
threshold \cite{vapnik_nature_2000}) given by
\begin{equation}
u_{i}=\Lambda w_{i}
\end{equation}
\begin{equation}
b=\Lambda V_{\mathrm{th}}
\end{equation}
where $\Lambda>0$ is chosen such that $\Lambda D=1$ (for either
$D_{\mathrm{in}}$ or $D_{\mathrm{out}}$). 

Together with the sign and norm constraints on the weights, $\boldsymbol{u}$
and $b$ must satisfy the linear constraints
\begin{align}
\forall\mu\ y^{\mu}\left(\boldsymbol{u}^{T}\boldsymbol{x}^{\mu}-b\right) & \ge1\label{eq:c1}\\
\forall i\ s_{i}u_{i} & \ge0\nonumber \\
b & \ge0\nonumber 
\end{align}
where $s_{i}=1$ if $w_{i}$ is excitatory and $s_{i}=-1$ if $w_{i}$
is inhibitory, and the quadratic constraint
\begin{equation}
\left|\boldsymbol{u}\right|^{2}\le b^{2}\Gamma^{2}/V_{\mathrm{th}}^{2},
\end{equation}
which enforces the constraint $\wnorm\le\Gamma$. 

For the effective weights and threshold, $\kappa_{\mathrm{in}}$ is
given by $\kappa_{\mathrm{in}}=\frac{1}{\left|\boldsymbol{u}\right|}$
and $\kappa_{\mathrm{out}}$ is given by $\kappa_{\mathrm{out}}=\frac{V_{\mathrm{th}}}{b}$.
Thus, maximizing $\kappa_{\mathrm{in}}$ is equivalent to minimizing
$\left|\boldsymbol{u}\right|$ and maximizing $\kappa_{\mathrm{out}}$
is equivalent to minimizing $b$. We therefor define a minimization
cost function $E\left(\boldsymbol{u},b\right)$ that is given by
\begin{equation}
E_{\text{in}}\left(\boldsymbol{u},b\right)=\frac{1}{2}\boldsymbol{u}^{T}\boldsymbol{u}\ ,
\end{equation}
for the $\kappa_{\mathrm{in}}^{\max}$ solution, and 
\begin{equation}
E_{\text{out}}\left(\boldsymbol{u},b\right)=b\ ,
\end{equation}
for the $\kappa_{\mathrm{out}}^{\max}$ solution.

To find the maximal $\kappa_{\mathrm{in}}$ or maximal $\kappa_{\mathrm{out}}$
solution we solve the conic program:
\begin{equation}
\min_{u,b,\tau}E\left(\boldsymbol{u},b\right)+\beta\tau\label{eq:opt_prob}
\end{equation}

in the limit of $\beta\rightarrow\infty$, subject to
\begin{align}
\forall\mu\ y^{\mu}\left(\boldsymbol{u}^{T}\boldsymbol{x}^{\mu}-b\right) & \ge1-\tau\label{eq:constraints}\\
\forall i\ s_{i}u_{i} & \ge0\nonumber \\
b & \ge0\nonumber \\
\tau & \ge0\nonumber \\
b^{2}\Gamma^{2}/V_{\mathrm{th}}^{2} & \ge\left|\boldsymbol{u}\right|^{2}\ .\nonumber 
\end{align}

$\tau$ is a global regularization variable that insures the existence
of a solution to the optimization problem (eqs. \ref{eq:opt_prob}
and \ref{eq:constraints}) even when the linear constraints (\ref{eq:c1})
are not realizable. In practice it is sufficient to set $\beta$ to
be a large constant (we set $\beta=10^{5}$). If the optimal value
of $\tau$ is zero the solution corresponds to the optimal perceptron
solution for the classification task. If the optimal value of $\tau$
is greater then zero, it indicates that the labeled patterns are not
linearly separable and that there is no zero error solution to the
classification task. Given that a solution with $\tau=0$ is found,
the optimal weights are given by \textbf{$\boldsymbol{w}=V_{\mathrm{th}}\boldsymbol{u}/b$.}

\section{\label{sec:Capacity-for-non-even}Capacity for non-even split of
`plus' and `minus' patterns}

The capacity of a perceptron with no sign constraints on synaptic
weights for classification of random patterns is a function of the
fraction of `plus' patterns in the desired classification, $p_{\mathrm{out}}$
\cite{gardner_maximum_1987,gardner_optimal_1988,gardner_space_1988}
and is given by 
\[
\alpha_{\mathrm{c}}^{\mathrm{Unconst.}}=\left[p_{\mathrm{out}}\int_{-\infty}^{\Delta}Dt\left(t-\Delta\right)^{2}+\left(1-p_{\mathrm{out}}\right)\int_{\Delta}^{\infty}Dt\left(t-\Delta\right)^{2}\right]^{-1}
\]
where $Dt$ is the Gaussian integration measure, $Dt=\frac{e^{-\frac{t^{2}}{2}}}{\sqrt{2\pi}}\mathrm{d}t$
and the order parameter $\Delta$ is given by the solution to the
equation
\[
0=p_{\mathrm{out}}\int_{-\infty}^{\Delta}Dt\left(t-\Delta\right)+\left(1-p_{\mathrm{out}}\right)\int_{\Delta}^{\infty}Dt\left(t-\Delta\right)\ .
\]
Fig. \ref{fig:res_1-1}(a) depicts the theoretical and measured $\alpha_{\mathrm{c}}$
of our `constrained' perceptron as a fraction of the corresponding
unconstrained capacity vs. $f_{\mathrm{exc}}$ for two values of $p_{\mathrm{out}}$.
Fig. \ref{fig:res_1-1}(b) depicts theoretical and measured of $\alpha_{\mathrm{b}}$
as a fraction of $\alpha_{\mathrm{c}}$ for two values of $p_{\mathrm{out}}$.

\section{Effects of excitatory and inhibitory input statistics}

Our results depend, of course, on parameters, but in a fairly reduced
way. In particular, the properties we discuss depend on the ratio
of the inputs standard deviations, $\lambda=\sigma_{\mathrm{inh}}/\sigma_{\mathrm{\ex}}$
and the ratio of their coefficients of variation, $\phi=\mathrm{CV_{\mathrm{exc}}/CV_{inh}}$
(see section \ref{sec:Replica-theoryr}). As discussed in the main
text $\phi$ determines the value of the optimal fraction of excitatory
synapses, $f_{\ex}^{\star}$ which can be written as $f_{\ex}^{\star}=\phi/\left(1+\phi\right)$
(see eq. 2 in the main text). Thus the shape of the phase diagram
changes with $\phi$ (Fig. \ref{fig:input_stats}a). The parameter
$\lambda$ has more subtle effects. We note here the main effect $\lambda$
has on the maximal $\kin$ and maximal $\kout$ solutions. 

\subsection{Balanced and unbalanced maximal $\protect\kin$ solutions}

The \textbf{maximal $\kin$ solutions} can be either balanced or unbalanced
depending on $f_{\ex},\ \phi\ ,\lambda$ and the value of $\kinmax$
(see example in Fig. \ref{fig:input_stats}b). Importantly, for a
wide range of reasonable parameters (For example, $\phi\le f_{\ex}/\left(1-f_{\ex}\right)$
and $\lambda\ge1$) the $\kinmax$ solution is unbalanced for all
values of $\kinmax$. 

\subsection{Fraction of `silent' weights in maximal $\protect\kin$ and maximal
$\protect\kout$ solutions}

As noted in previous studies \cite{brunel_optimal_2004,chapeton_efficient_2012},
a prominent feature of `critical' solutions with sign constraint weights,
such as the maximal $\kin$ and maximal $\kout$ solutions, is that
a finite fraction of the synapses are `silent' i.e. $w_{i}=0$. Our
theory allows us to derive the full distribution of synaptic efficacies
(section \ref{subsec:Distribution-of-synaptic}) and calculate the
fraction of silent weights for each solution. For the maximal $\kout$
solutions in the unbalanced regime ($\alpha_{\mathrm{b}}<\alpha<\alpha_{\mathrm{c}}$),
the fraction of excitatory (inhibitory) silent weights is always larger
(smaller) then 1/2 (Fig. \ref{fig:input_stats}c). However, in the
balanced regime ($\alpha<\alpha_{\mathrm{b}}$) the qualitative behavior
depends on $\lambda$ (see Fig. \ref{fig:input_stats}c). Interestingly,
for \textbf{unbalanced maximal $\kin$ solutions} the fraction of
silent weights is constant and equals 1/2 for both excitatory and
inhibitory inputs (see sections \ref{subsec:max_kappa_in} and \ref{subsec:Distribution-of-synaptic}).

\subsection{Tuning properties of cortical neurons suggest that in cortex \textmd{\normalsize{}$f_{\protect\ex}^{\star}>0.5$}}

In cortical circuits, inhibitory neurons tend to fire with higher
firing rates and are thought to be more broadly tuned than excitatory
neurons implying, under reasonable assumptions, that both $\lambda$
and $\phi$ are greater than 1 leading to $f_{\ex}^{\star}>0.5$.

To see this, we consider input neurons with Gaussian tuning curves
to some external stimulus variable $\varphi\in\left[0,1\right]$,
i.e. the mean response, $x_{i}$, of neuron $i$ to stimulus $\varphi$
is given by 
\begin{equation}
x_{i}=A_{i}\exp\left[-\left(\varphi-\varphi{}_{i}^{\mathrm{pref}}\right)^{2}/\left(2\delta_{i}^{2}\right)\right]\ ,
\end{equation}
where $A_{i}$, $\varphi_{i}^{\mathrm{pref}}$ and $\delta_{i}$ characterize
the response properties of the neuron. Assuming that $\varphi$ is
distributed uniformly, and, for simplicity, that $\delta_{i}\ll1$,
the mean and variance of the neurons' responses are given by:
\begin{equation}
\bar{x}_{i}=\sqrt{2\pi}A_{i}\delta_{i}\ ,
\end{equation}
and
\begin{equation}
\sigma_{i}^{2}\simeq A^{2}\delta\sqrt{\pi}\ ,
\end{equation}
where we neglect terms of order $\delta^{2}$. We now assume that
$A_{i}=A_{\ex}$ and $\delta_{i}=\delta_{\ex}$ if neuron $i$ is
excitatory and that $A_{i}=A_{\inh}$ and $\delta_{i}=\delta_{\inh}$
if neuron $i$ is inhibitory. Further, we assume that inhibitory neurons
respond with a higher firing rate ($A_{\inh}>A_{\ex}$) and are more
broadly tuned ($\delta_{\inh}>\delta_{\ex})$. In this case we have
\begin{equation}
\lambda=\frac{A_{\inh}\sqrt{\delta_{\inh}}}{A_{\ex}\sqrt{\delta_{\ex}}}>1\ ,
\end{equation}
and 
\begin{equation}
\phi=\sqrt{\frac{\delta_{\inh}}{\delta_{\ex}}}>1\ .
\end{equation}

\section{$\protect\koutmax$ and $\protect\kinmax$ solutions in purely excitatory
networks}

In purely excitatory networks ($f_{\ex}=1$) all solutions are unbalanced
and output robustness can be achieved by sparse input \cite{brunel_optimal_2004}
or tonic inhibition \cite{brunel_is_2016}. However, the distinction
between output robustness and input robustness still apply and, surprisingly,
maximizing either $\kin$ or $\kout$ leads to two different solutions
with qualitatively different properties.

In particular, as noted in \cite{brunel_optimal_2004,chapeton_efficient_2012},
the fraction of silent weights of the $\koutmax$ solutions increases
as the load decreases. Thus, if the network implements the maximal
$\kout$ solution, network connectivity, as measured in pairwise stimulation
experiment, is expected to be sparse. However, for the maximal $\kin$
solution the fraction of silent weights is constant and remains $1/2$
for all values of the load. Thus, measured network connectivity is
expected to be higher. 

Establishing correspondence between theory and experiment in this
case is confounded by the difficulty to experimentally distinguish
between silent synapses and completely absent synapses that were never
available as inputs for the post synaptic neuron during learning. 

\section{\label{sec:non_learned_inh}Recurrent networks with non-learned inhibition}

In our basic model for an associative memory network we assume that
the activity of both excitatory and inhibitory neuron is specified
in the desired memory states and that all network connections are
learned. Both of these assumptions can be modified creating new scenarios
with different computational properties.

First we assume that memory state is only specified by the activity
of excitatory neurons and that the memory is recalled when the activity
of excitatory neurons matches the memory state regardless of the activity
of inhibitory neurons. The problem of learning in such a network is
computationally hard since the learning needs to optimize the activity
of the inhibitory neurons using the full connectivity matrix. In our
work we do not address this scenario. Instead we forgo the assumption
that excitatory and inhibitory connections onto inhibitory neurons
are learned and replace them with randomly chosen connections, i.e.
assume that E to I and I to I connections are not learned and random.

\subsection{Choosing random synapses for inhibitory neurons}

In this scenario the activity of inhibitory neurons is determined
by the network dynamics. We consider random I to I and E to I weights
with means $J_{\mathrm{II}}$ and $J_{\mathrm{IE}}$ and standard
deviations $\sigma_{J_{\mathrm{II}}}$ and $\sigma_{J_{\mathrm{IE}}}$.
We will examine the distribution of inhibitory neurons' membrane potential
given that the activity of excitatory neurons is held at a memory
state in which $p_{\mathrm{out}}^{\mathrm{exc}}N$ neurons are active.
When $N$ is large, this distribution is Gaussian and we assume correlations
are weak. Thus, the mean activity in the network is the probability
that the membrane potential is above threshold and is given by the
equation
\begin{equation}
m_{\mathrm{I}}=H\left(\frac{V_{\mathrm{th}}-\left\langle V\right\rangle }{\sqrt{\sigma^{2}\left(V\right)}}\right)\ ,\label{eq:mean_inh}
\end{equation}
where $H\left(x\right)=\int_{x}^{\infty}\frac{e^{-\frac{y^{2}}{2}}}{\sqrt{2\pi}}\mathrm{d}y$,
and $\left\langle V\right\rangle $ and $\sigma^{2}\left(V\right)$
are the mean and variance of the membrane potential of inhibitory
neurons, respectively. 

On the other hand, given the mean activity, $m_{\mathrm{I}}$, the
mean and variance of the membrane potentials are given by 
\begin{equation}
\left\langle V\right\rangle =N\left(p_{\mathrm{out}}^{\mathrm{exc}}g_{\mathrm{exc}}J_{\mathrm{IE}}-m_{\mathrm{I}}\left(1-g_{\mathrm{exc}}\right)J_{\mathrm{II}}\right)
\end{equation}
\begin{equation}
\sigma^{2}\left(V\right)=N\left(\left[\sigma_{J_{\mathrm{IE}}}^{2}+\left(1-p_{\mathrm{out}}^{\mathrm{exc}}\right)J_{\mathrm{IE}}^{2}\right]p_{\mathrm{out}}^{\mathrm{exc}}g_{\mathrm{exc}}+\left[\sigma_{J_{\mathrm{II}}}^{2}+\left(1-m_{\mathrm{I}}\right)J_{\mathrm{II}}^{2}\right]m_{\mathrm{I}}\left(1-g_{\mathrm{exc}}\right)\right)\ .\label{eq:var_V}
\end{equation}
Together, eqs. (\ref{eq:mean_inh})-(\ref{eq:var_V}) define the relations
between $m_{\mathrm{I}}$, $J_{\mathrm{II}}$ , $J_{\mathrm{IE}}$,
$\sigma_{J_{\mathrm{IE}}}$ and $\sigma_{J_{\mathrm{II}}}$. 

In our simulations we set $m_{\mathrm{I}}$ and the mean and variance
of the I to I connections, and choose the mean and variance of the
E to I connections according to the solution of (\ref{eq:mean_inh})
(When $N$ is large $J_{\mathrm{IE}}$ is given by $NJ_{\mathrm{IE}}\simeq\left(V_{\mathrm{th}}+m_{\mathrm{I}}N\left(1-g_{\mathrm{exc}}\right)J_{\mathrm{II}}\right)/\left(g_{\mathrm{exc}}p_{\mathrm{out}}^{\mathrm{exc}}\right)$).
In particular, we choose an inhibitory network with binary weights
in which each inhibitory neuron projects to another inhibitory neuron
with probability $p_{\mathrm{II}}$ with synaptic efficacy $j_{\mathrm{II}}=1/\left(\sqrt{N}p_{\mathrm{II}}\right)$.
Each excitatory neuron project to an inhibitory neuron with probability
$p_{\mathrm{IE}}$ with synaptic efficacy $j_{\mathrm{IE}}$ that
ensures that the mean inhibitory activity level at the memory states
is $m_{\mathrm{I}}$.

In this parameter regime, the inhibitory subnetwork exhibits asynchronous
activity, with mean activity $m_{\mathrm{I}}$, at the excitatory
memory states. However, different memory states lead to different
asynchronous states. 

\subsection{Training set definition}

Excitatory neurons need to learn to remain stationary at the desired
memory states given the network activity at this state. However, since
the activity of the inhibitory subnetwork is not stationary at the
desired memory states, the training set for learning is not well defined. 

To properly define the training set, we sample $n_{\mathrm{sample}}$
instances of the generated inhibitory activity for each memory state
when the activity of the excitatory neuron is clipped to this memory
state. Sampling was performed by running the inhibitory network dynamics
and recording the state of the inhibitory neurons after $T=100N$
time steps. We then use the sampled activity patterns together with
the excitatory memory states as an extended training set (with $Pn_{\mathrm{sample}}$
patterns) for the excitatory neurons.

\subsection{Learned network stability}

The non fixed point dynamics of the inhibitory subnetwork implies
that the convergence of the learning on the training set does not
entail that the memory states themselves are dynamically stable, in
contrast to our prior model in which inhibitory neurons learn their
synaptic weights. Therefor, after training we measure the probabilities
that patterns are stable. This is done by the following procedure:
First we run the network dynamics (with $\sigma_{\mathrm{out}}=0$)
when the excitatory neurons' activity is clipped to the memory state,
for $T_{\mathrm{init}}=50N$ time steps. We then release the excitatory
neurons to evolve according to the natural network dynamics and observe
if their activity remain in the vicinity of the memory state for $T=500N$
time steps. In a similar way we test the basins of attractions, starting
the excitatory network from a distorted version of the memory state
instead of the memory state itself. 

\subsection{Learning only E to E connections}

First we consider the case in which I to E connections are random:
Each inhibitory neuron project to an excitatory neuron with probability
$p_{\mathrm{EI}}$ with synaptic efficacy $j_{\mathrm{EI}}=1/\left(\sqrt{N}p_{\mathrm{EI}}\right)$.
We then try to find appropriate E to E connection using the learning
scheme described above. We find that the pattern to pattern fluctuations
in the inhibitory feedback due to the variance of the I to E connections,
and the variance in the inhibitory network neurons' activation is
substantial and of the same order of the signal differentiating the
memory states. In fact, in this scenario the parameters we consider
($N=2000$, $P=1000,$ $g_{\mathrm{exc}}=0.8$, $p_{\mathrm{out}}^{\mathrm{exc}}=0.15,$
$p_{\mathrm{II}}=1/2,$ $p\mathrm{_{IE}=1/2},$ $m_{\mathrm{I}}=0.4$,
$p_{\mathrm{EI}}=1/2$, $n_{\mathrm{sample}}=40$) are above the system\textquoteright s
memory capacity and we are unable to find appropriate excitatory weights
which implement the desired memory states for the training set. We
conclude that this form of balancing inhibitory feedback is too restrictive
due to the heterogeneity of I to E connections and variability of
inhibitory neurons' activity. 

\subsection{Learning both E to E and I to E connections}

In this scenario we find the maximal $\kappa_{\mathrm{out}}$ solution
for the extended training set described above. For the parameters
used ($N=2000$, $P=1000,$ $g_{\mathrm{exc}}=0.8$, $p_{\mathrm{out}}^{\mathrm{exc}}=0.15,$
$p_{\mathrm{II}}=1/2,$ $p\mathrm{_{IE}=1/2},$ $m_{\mathrm{I}}=0.4,$
$n_{\mathrm{sample}}=40$) we are able to find solutions that implement
all the desired memory states for the extended training set. In addition,
we find that the excitatory memory states are dynamically stable with
very high probability (we did not observe any unstable pattern). For
numerical results see Fig. 5 and Fig \ref{fig:Effect-of-shuffling}. 

\section{\label{sec:Replica-theoryr}Replica theory for sign and norm constrained
perceptron}

We use the Replica method \cite{engel_statistical_2001} to calculate
the system's typical properties. For the Perceptron architecture the
replica symmetric solution has been shown to be stable and exact \cite{gardner_maximum_1987,gardner_optimal_1988,gardner_space_1988}.

Given a set of $P$ patterns, $\boldsymbol{x}^{\mu},$ and desired
labels $y^{\mu}=\pm1$ for $\mu=1,2,\dots,P$, the Gardner volume
for is given by:
\begin{equation}
V_{\mathrm{G}}=\int\mathrm{\mathcal{D}\left(\boldsymbol{w}\right)}\prod_{\mu=1}^{P}\Theta\left[y^{\mu}\left(w^{\mathrm{T}}\boldsymbol{x}^{\mu}-V_{\mathrm{th}}\right)-K\right]\ ,
\end{equation}
where $\Theta\left[x\right]$ is the Heaviside step function and $\mathrm{\mathcal{D}\left(\boldsymbol{w}\right)}$
is an integration domain obeying the sign and norm constraint $\wnorm\le\Gamma$.

We assume input pattern and labels are drawn independently from distribution
with non negative means $\bar{x}_{\ex\left(\inh\right)}$ and standard
deviation $\sigma_{\ex\left(\inh\right)}$. Labels are independently
drawn from a binary distribution with $\mathrm{Pr}\left(y^{\mu}=1\right)=p_{\mathrm{out}}$
and $\mathrm{Pr}\left(y^{\mu}=-1\right)=1-p_{\mathrm{out}}$.

We handle both input and output robustness criteria by using different
$K$ for each case:
\begin{align}
K_{\text{in}} & =\wnorm\sigma_{\mathrm{exc}}\kappa_{\mathrm{in}}\\
K_{\text{out}} & =V_{\mathrm{th}}\kappa_{\mathrm{out}}\ ,\nonumber 
\end{align}
where, here, $\kappa_{\mathrm{in}}$ and $\kout$ are dimensionless
numbers representing the input robustness in units of $\sigma_{\mathrm{exc}}$
and the output robustness in units of $V_{\mathrm{th}}$ respectively. 

Further, we define the parameters:
\begin{equation}
\lambda=\frac{\sigma_{\mathrm{inh}}}{\sigma_{\mathrm{exc}}},\ \eta=\frac{\bar{x}_{\mathrm{inh}}}{\bar{x}_{\mathrm{exc}}}\ ,
\end{equation}
 and 
\begin{equation}
\phi=\frac{\mathrm{CV_{exc}}}{\mathrm{CV_{inh}}}\ .
\end{equation}

\subsection{The order parameters}

We calculate the mean logarithm of the Gardner volume $\left\langle \left\langle \ln V_{\mathrm{G}}\right\rangle _{x}\right\rangle _{y}$
averaged over the excitatory and inhibitory input distributions, and
the desired label distribution. The result of the calculation is expressing
$\left\langle \left\langle \ln V_{\mathrm{G}}\right\rangle _{x}\right\rangle _{y}$
as a stationary phase integral over a free energy that is a function
of several order parameters. The value of the order parameters is
determined by the saddle point equations of the free energy.

In our model the saddle point equations are a system of six equations
for the six order parameter: $q,\ Q,\ \theta,\ \Delta,\ B$ and $C$. 

The order parameters $q,\ Q,\ \theta,\ \mathrm{and}\ \Delta\ $ have
a straight forward physical interpretation. 

The parameter $q$ is the mean typical correlation coefficient between
the $V_{\mathrm{PSP}}$'s elicited by two different solutions to the
same classification task: given two typical solution weight vectors
$\boldsymbol{w}^{\alpha}$ and $\boldsymbol{w}^{\beta}$, $q$ is
given by 
\begin{equation}
q=\frac{\sum_{i=1}^{N}\lambda_{i}^{2}w_{i}^{\alpha}w_{i}^{\beta}}{\sqrt{\left(\sum_{i=1}^{N}\lambda_{i}^{2}w_{i}^{\alpha2}\right)\left(\sum_{i=1}^{N}\lambda_{i}^{2}w_{i}^{\beta2}\right)}}\ ,
\end{equation}
where $\lambda_{i}=1$ if $w_{i}$ is excitatory and $\lambda_{i}=\lambda$
if $w_{i}$ is inhibitory. 

Given a typical solution $\boldsymbol{w}$, the physical interpretation
of $Q$ and $\theta$ is given by 
\begin{equation}
Q=\frac{\sum_{i=1}^{N}\lambda_{i}^{2}w_{i}^{2}}{\sum_{i=1}^{N}w_{i}^{2}}
\end{equation}
\begin{equation}
\theta=\frac{V_{\mathrm{th}}}{\sigma_{\ex}}\left(\sum_{i=1}^{N}\lambda_{i}^{2}w_{i}^{2}\right)^{-\frac{1}{2}}\:.
\end{equation}

The norm constraint on the weights is satisfied as long as 
\begin{equation}
\theta\ge\frac{V_{\mathrm{th}}}{\sigma_{\ex}\Gamma\sqrt{Q}}\ .\label{eq:theta_lower_bound-1-1}
\end{equation}
Thus, there are two types of solutions. One in which the value of
$\theta$ is determined by the saddle point equation (unbalanced solutions)
and the other in which $\theta$ is clipped to its lower bound value
(balanced solutions). Notice that $q$ and $Q$ remain of order 1
for any scaling of $\wnorm$ while $\theta$ scales as $\sqrt{N}$
when $\wnorm$ is of order $1/\sqrt{N}$ and is of order $1$ when
$\wnorm$ is of order 1. 

The physical interpretation of $\Delta$ can be expressed through
the following relation 
\begin{equation}
\Delta=\theta\left(1-\frac{\bar{x}_{\ex}\sum_{i=1}^{N}\eta_{i}w_{i}}{V_{\mathrm{th}}}\right)\ ,
\end{equation}
where $\eta_{i}=1$ if $w_{i}$ is excitatory and $\eta_{i}=\eta$
if $w_{i}$ is inhibitory.

\subsection{Summary of main results}

Before describing the full saddle point (SP) equations and their various
solutions in detail we will provide a brief general summary of the
results that would hopefully provide some flavor of the derivations
for the interested reader. 

Since $\theta$ is bounded from bellow by $V_{\mathrm{th}}/\left(\sigma_{\ex}\Gamma\sqrt{Q}\right)$
we have two sets of SP equations which we term the balanced and the
unbalanced sets. In both sets, given the free energy $\mathcal{F}\left(Q,q,\Delta,\theta,B,C\right)$,
five of the SP equations are given by 
\begin{equation}
\frac{\partial\mathcal{F}}{\partial Q}=\frac{\partial\mathcal{F}}{\partial q}=\frac{\partial\mathcal{F}}{\partial\Delta}=\frac{\partial\mathcal{F}}{\partial B}=\frac{\partial\mathcal{F}}{\partial C}=0\ .
\end{equation}
The sixth equations is 
\begin{equation}
\frac{\partial\mathcal{F}}{\partial\theta}=0\label{eq:t1}
\end{equation}
in the unbalanced set and is 
\begin{equation}
\theta=\frac{V_{\mathrm{th}}}{\sigma_{\ex}\Gamma\sqrt{Q}}\label{eq:t2}
\end{equation}
in the balanced set. Importantly, we find that eq. \ref{eq:t1} has
solutions only when $\theta\sim\sqrt{N}$ which implies $\wnorm\sim1/\sqrt{N}$,
and eq. \ref{eq:t2} implies $\wnorm=\Gamma\sim\mathcal{O}\left(1\right)$,
justifying the naming of the two sets. The solutions to the two sets
of SP equations define the range of possible values of $\alpha$,
$\kappa_{\mathrm{in}}$ and $\kappa_{\mathrm{out}}$ that permits
the existence of solution weight vectors. There are a number of interesting
cases that we analyze below.

We first consider the solutions of the SP equations for zero $\kout$
and $\kin$. In this case the SP describes the typical solutions that
dominate the Gardner volume. Since the $N$-dim. volume of balanced
solutions with $\wnorm\sim1$ is exponentially larger than the volume
of unbalanced solutions with $\wnorm\sim1/\sqrt{N}$ we expect that
balanced solutions will dominate the Gardner volume whenever they
exist. Indeed, solving the two sets of SP equations we find that solutions
to the balanced set exist only for $\alpha<\alpha_{\mathrm{b}}$ while
solutions for the unbalanced set exist only for $\alpha_{\mathrm{b}}<\alpha<\alpha_{\mathrm{c}}$. 

Next, we examine the values of $\kout$ that permits solutions to
the balanced and unbalanced sets of SP equations. Importantly, we
show that the unbalanced set can only be solved for $\kout\propto1/\sqrt{N}$.
Thus, unbalanced solutions can not have $\kout$ of $\mathcal{O}\left(1\right)$
or conversely all solutions with $\kout$ of $\mathcal{O}\left(1\right)$
are balanced. 

Of particular interest are the so called `critical' solutions for
which $q\rightarrow1$. At this limit the typical correlation coefficient
between the $V_{\mathrm{PSP}}$'s elicited by two different solutions
to the same classification task approaches unity, which implies that
only one solution exist and the Gardner volume shrinks to zero. Thus,
for a given $\kappa_{\mathrm{in}}$ or $\kappa_{\mathrm{out}}$, the
value of $\alpha$ for which $q\rightarrow1$ is the maximal load
for which solutions exist. In this case, the SP describes the properties
of the maximal $\kout$ or $\kin$ solutions. 

The structure of the equations in this limit is relatively simple.
First, the order parameter $\Delta$ is given by the solutions to
\begin{equation}
0=p_{\mathrm{out}}\int_{-\infty}^{\Delta+\tilde{K}}Dt\left(t-\Delta-\tilde{K}\right)+\left(1-p_{\mathrm{out}}\right)\int_{\Delta-\tilde{K}}^{\infty}Dt\left(t-\Delta+\tilde{K}\right)\label{eq:D1}
\end{equation}
with the robustness parameter $\tilde{K}$ being $\tilde{K}_{\text{in}}=\kappa_{\mathrm{in}}/\sqrt{Q}$
or $\tilde{K}_{\text{out}}=\theta\kappa_{\mathrm{out}}$ and the integration
measure, $Dt$, is given by $Dt=\frac{e^{-\frac{t^{2}}{2}}}{\sqrt{2\pi}}\mathrm{d}t$.
Second, we find a simple relation between critical loads of the constrained
perceptron considered here and the critical loads of the classic unconstrained
perceptron, $\alpha^{\mathrm{Unconst.}}$:
\begin{equation}
\alpha=2C\alpha^{\mathrm{Unconst.}}\ .\label{eq:a}
\end{equation}
$\alpha^{\mathrm{Unconst.}}$ is given by 
\begin{equation}
\alpha^{\mathrm{Unconst.}}=\left[p_{\mathrm{out}}\int_{-\infty}^{\Delta+\tilde{K}}Dt\left(t-\Delta-\tilde{K}\right)^{2}+\left(1-p_{\mathrm{out}}\right)\int_{\Delta-\tilde{K}}^{\infty}Dt\left(t-\Delta+\tilde{K}\right)^{2}\right]^{-1}\!,\label{eq:a2}
\end{equation}
which is indeed the critical load of an unconstrained perceptron with
a given margin $\tilde{K}$ (see \cite{gardner_space_1988}). Finding
the critical load is then reduced to solving for the order parameter
$C$. For each value of $\kin>0$ or $\kout>0$ only one set of SP
equations can be solved, determining if the maximal $\kin$ or $\kout$
solutions are balanced or unbalanced. By examining the range of solutions
for each set we can find the value of $\kinmax$ and $\koutmax$ for
any $\alpha$ and determine that (a) the maximal $\kout$ solution
is balanced for $\alpha\le\alpha_{\mathrm{b}}$ and unbalanced for
$\alpha_{\mathrm{b}}<\alpha<\mathrm{\alpha_{c}}$ and (b) that for
a wide range of parameters the maximal $\kin$ is unbalanced for all
$\alpha<\alpha_{\mathrm{c}}$. In addition, we find that for $\alpha_{\mathrm{b}}<\alpha<\mathrm{\alpha_{c}}$
, $\koutmax$ is given by 
\begin{equation}
\koutmax=\frac{\sigma_{\ex}}{\bar{x}_{\ex}\sqrt{N}}\kappa_{0}
\end{equation}
where $\kappa_{0}$ is finite and larger than zero when $\alpha$
approaches $\alpha_{\mathrm{b}}$ from above and $\kappa_{0}$ approaches
zero when $\alpha$ approaches $\alpha_{\mathrm{c}}$. The above result
implies that output robustness can be increased when the tuning of
the input is increased. As we discuss in the main text in the context
of neuronal selectivity in purely excitatory circuits, sparse input
activity is one way to increase the input tuning. If we consider sparse
binary inputs with mean activity level $s\ll1$ the output robustness
will be given by $\kappa_{0}\sqrt{\left(1-s\right)/sN}\simeq\kappa_{0}/\sqrt{sN}$. 

Finally, we consider the solutions of the SP equations in the critical
limit ($q\rightarrow1$) for $\kin=\kout=0$. In this limit the SP
describes the capacity and balanced capacity. We note that for $\kout=\text{\ensuremath{\kin}=}0$,
$\Delta$ (and, as a result $\alpha_{\mathrm{c}}^{\mathrm{Unconst.}}$)
is independent of all the other order parameters, simplifying the
equations. In this case, we have only two coupled SP equations (for
the order parameters $B$, $C$ and $\theta$), given by 
\begin{align}
\left(1-\frac{\sigma_{\ex}B\theta}{\bar{x}_{\ex}\sqrt{2CN}}\right)^{2}C & =f_{\mathrm{exc}}\gamma_{+}\left(B\right)+\left(1-f_{\mathrm{exc}}\right)\gamma_{-}\left(B\phi\right)\label{eq:C_critical-1}\\
\frac{\sigma_{\ex}\theta}{\bar{x}_{\ex}\sqrt{N}} & =-\frac{f_{\mathrm{exc}}\gamma_{+}^{\prime}\left(B\right)+\left(1-f_{\mathrm{exc}}\right)\phi\gamma_{-}^{\prime}\left(B\phi\right)}{\sqrt{2C}}
\end{align}
where we defined the functions:
\begin{align}
\gamma_{\pm}\left(x\right) & =\frac{x^{2}+1}{2}-\frac{1}{2}\int Du\left(x+u\right)^{2}\Theta\left[\pm\left(x+u\right)\right]\\
\gamma_{\pm}^{\prime}\left(x\right) & =x-\int Du\left(x+u\right)\Theta\left[\pm\left(x+u\right)\right]\ .
\end{align}
For the balanced SP equations we have $\theta=V_{\mathrm{th}}/\left(\sigma_{\ex}\Gamma\sqrt{Q}\right)$
and for the unbalanced SP equations, eq. \ref{eq:t1} reduces to $B=0$.
Finally, $\alpha$ is given by (\ref{eq:a}). 

For the unbalanced set ($B=0$) we have $\gamma_{\pm}\left(0\right)=\frac{1}{2}-\frac{1}{2}\int_{0}^{\infty}Du\left(u\right)^{2}=\frac{1}{4}$
and $\gamma_{\pm}^{\prime}\left(0\right)=\mp\int_{0}^{\infty}Du\left(u\right)=\mp\frac{1}{\sqrt{\pi}}$.
We immediately get  $C=\frac{1}{4}$ and 
\begin{equation}
\theta=\sqrt{\frac{2N}{\pi}}\left[f_{\mathrm{exc}}/\mathrm{CV_{\mathrm{exc}}-\left(1-f_{\mathrm{exc}}\right)/}\mathrm{CV}_{\mathrm{inh}}\right]\ .
\end{equation}
This solution suggests that at capacity the solutions are unbalanced
$\left(\theta\sim\sqrt{N}\Rightarrow\wnorm\sim1/\sqrt{N}\right)$
and that capacity as a function of $f_{\mathrm{exc}}$ is constant
with
\begin{equation}
\alpha_{\mathrm{c}}=\frac{1}{2}\alpha_{\mathrm{c}}^{\mathrm{Unconst.}}\ .
\end{equation}
However, this solution is only valid as long as $\theta>V_{\mathrm{th}}/\left(\sigma_{\ex}\Gamma\sqrt{Q}\right)$
which is true only as long as 
\begin{equation}
\left[f_{\mathrm{exc}}/\mathrm{CV_{\mathrm{exc}}}-\left(1-f_{\mathrm{exc}}\right)/\mathrm{CV}_{\mathrm{inh}}\right]>0
\end{equation}
 which implies 
\begin{equation}
f_{\mathrm{exc}}>f_{\mathrm{exc}}^{\star}=\frac{\mathrm{CV_{\mathrm{exc}}}}{\mathrm{CV}_{\ex}+\mathrm{CV}_{\mathrm{inh}}}\ .
\end{equation}
For the solution for the balanced set ($\theta=V_{\mathrm{th}}/\left(\sigma_{\ex}\Gamma\sqrt{Q}\right)$)
terms with $\theta/\sqrt{N}$ can be neglected and we have the equation
\begin{equation}
0=f_{\mathrm{exc}}\gamma_{+}^{\prime}\left(B\right)+\left(1-f_{\mathrm{exc}}\right)\phi\gamma_{-}^{\prime}\left(B\phi\right)\label{eq:b1}
\end{equation}
 for the order parameter $B$. $C$ and $Q$ are given by 
\begin{align}
C & =f_{\mathrm{exc}}\gamma_{+}\left(B\right)+\left(1-f_{\mathrm{exc}}\right)\gamma_{-}\left(B\phi\right)\label{eq:C_critical-1-1}\\
Q & =\frac{f_{\mathrm{exc}}\gamma_{+}\left(B\right)+\left(1-f_{\mathrm{exc}}\right)\gamma_{-}\left(B\phi\right)}{f_{\mathrm{exc}}\gamma_{+}\left(B\right)+\left(\frac{1-f_{\mathrm{exc}}}{\lambda^{2}}\right)\gamma_{-}\left(B\phi\right)}
\end{align}
This solution gives us the balanced capacity line
\begin{equation}
\alpha_{\mathrm{b}}\left(f_{\mathrm{exc}}\right)=2\left[f_{\mathrm{exc}}\gamma_{+}\left(B\right)+\left(1-f_{\mathrm{exc}}\right)\gamma_{-}\left(B\phi\right)\right]\alpha_{\mathrm{}}^{\mathrm{Uncont.}}\left(p_{\mathrm{out}}\right)
\end{equation}
where $B$ is given by the solution of (\ref{eq:b1}) and $\alpha_{\mathrm{}}^{\mathrm{Uncont.}}\left(p_{\mathrm{out}}\right)$
is given by (\ref{eq:a2}) and (\ref{eq:D1}) with $\tilde{K}=0$.

\subsection{Detailed solutions of the saddle point equations}

Below we provide the saddle point equations and their solutions under
various conditions. We also provide the derived form of the distributions
of synaptic weights for critical solutions.

\subsubsection{The general saddle point equations}

We define the following:
\begin{align}
\mathcal{F}_{h}= & p_{\mathrm{out}}\int Dt\ln H\left[-X_{+}\left(t\right)\right]+\left(1-p_{\mathrm{out}}\right)\int Dt\ln H\left[X_{-}\left(t\right)\right]\\
Dt & =\frac{e^{-\frac{t^{2}}{2}}}{\sqrt{2\pi}}\mathrm{d}t\\
H\left(x\right) & =\int_{x}^{\infty}Dt\\
X_{\pm}\left(t\right)= & \frac{\sqrt{q}t-\Delta\mp\tilde{K}}{\sqrt{1-q}}\\
\tilde{K}_{\text{in}} & =\kappa_{\mathrm{in}}/\sqrt{Q}\\
\tilde{K}_{\text{out}} & =\theta\kappa_{\mathrm{out}}\\
\phi_{+} & =1,\ \phi_{-}=\phi\\
\lambda_{+} & =1,\ \lambda_{-}=\lambda\\
f_{+} & =f_{\ex},\ f_{-}=1-f_{\ex}\\
\alpha & =\frac{P}{N}\\
\tilde{\theta} & =\frac{\sigma_{\ex}\left(\theta-\Delta\right)}{\bar{x}_{\ex}\sqrt{N}}\\
Z_{\pm}= & 2C\left[2C-\sqrt{2C}B\tilde{\theta}+\left(1-q\right)\left(1-2\alpha Q\frac{\partial\mathcal{F}_{h}}{\partial Q}\left(1-\frac{Q}{\lambda_{\pm}^{2}}\right)\right)\right]^{-1}\\
\Phi_{\pm}\left(x,z,q\right) & =\frac{z\left(x^{2}+1\right)}{2}+\frac{1-q}{2}\left[1+\int DuJ_{1}\left(\pm\frac{\sqrt{z}\left(x+u\right)}{\sqrt{\left(1-q\right)}}\right)\right]\\
\Phi_{\pm}^{\prime}\left(x,z,q\right) & =zx\pm\sqrt{z\left(1-q\right)}\int DuJ_{2}\left(\pm\frac{\sqrt{z}\left(x+u\right)}{\sqrt{\left(1-q\right)}}\right)\\
J_{1}\left(x\right)= & \frac{H^{\prime}\left(x\right)}{H\left(x\right)}x\\
J_{2}\left(x\right)= & \frac{H^{\prime}\left(x\right)}{H\left(x\right)}
\end{align}
The saddle point equations are given by:
\begin{align}
C/Q & =\sum_{\pm}\frac{f_{\pm}}{\lambda_{\pm}^{2}}Z_{\pm}\Phi_{\pm}\left(B\phi,Z_{\pm},q\right)\label{eq:w_norm_eq-1-1}\\
C & =\sum_{\pm}f_{\pm}Z_{\pm}\Phi_{\pm}\left(B\phi_{\pm},Z_{\pm},q\right)\\
\sqrt{2C}\tilde{\theta} & =-\sum_{\pm}f_{\pm}\phi_{\pm}\Phi_{\pm}^{\prime}\left(B\phi_{\pm},Z_{\pm},q\right)\\
\alpha & =-\frac{C}{\left(1-q\right)^{2}}\left(\frac{\partial\mathcal{F}_{h}}{\partial q}\right)^{-1}\label{eq:alpha_t}\\
\frac{\sigma_{\mathrm{exc}}B}{\bar{x}_{\mathrm{exc}}\sqrt{2CN}} & =-\frac{1}{2\left(1-q\right)}\frac{\partial\mathcal{F}_{h}}{\partial\Delta}\left(\frac{\partial\mathcal{F}_{h}}{\partial q}\right)^{-1}\\
\frac{\sigma_{\mathrm{exc}}B}{\bar{x}_{\mathrm{exc}}\sqrt{2CN}}= & \frac{1-q}{2\theta C}+\frac{1}{2\left(1-q\right)}\frac{\partial\mathcal{F}_{h}}{\partial\theta}\left(\frac{\partial\mathcal{F}_{h}}{\partial q}\right)^{-1}\ \mathrm{\boldsymbol{OR}}\ \theta=\frac{V_{\mathrm{th}}}{\sigma_{\ex}\Gamma\sqrt{Q}}
\end{align}
It is important to note the relation between $\theta$ and $\tilde{\theta}$.
$\tilde{\theta}$ is of the order of $\theta/\sqrt{N}$ ($\Delta$
remains of order 1 under all conditions). Thus, for unbalanced solutions
$\theta\sim\sqrt{N}$ and $\tilde{\theta}$ is of $\order 1$ while
for balanced solutions $\theta$ is of $\order 1$ and $\tilde{\theta}$
is of $\order{1/\sqrt{N}}$ and can be neglected. 

\subsubsection{Saddle point equations for typical solutions}

For typical solutions we solve the saddle point equations for $\kin=0$
or $\kout=0$ leading to $K=0$. In this case we have $\frac{\partial\mathcal{F}_{h}}{\partial\theta}=\frac{\partial\mathcal{F}_{h}}{\partial Q}=0$
and thus 
\begin{equation}
Z_{\pm}=2C\left[2C-\sqrt{2C}B\tilde{\theta}+\left(1-q\right)\right]^{-1}
\end{equation}
and the saddle point equation for $\theta$ is 
\begin{equation}
\frac{\sigma_{\mathrm{exc}}\sqrt{2C}B\theta}{\bar{x}_{\mathrm{exc}}\sqrt{N}}=1-q\ \mathrm{\boldsymbol{OR}}\ \theta=\frac{V_{\mathrm{th}}}{\sigma_{\ex}\Gamma\sqrt{Q}}
\end{equation}
We now can solve the saddle point equations for the unbalanced case
with:
\begin{align}
Z_{\pm}=1\  & \mathrm{and}\ \theta>\frac{V_{\mathrm{th}}}{\sigma_{\ex}\Gamma\sqrt{Q}},\ \tilde{\theta}>0
\end{align}
and for the balanced case with 
\begin{equation}
Z_{\pm}=2C\left[2C+\left(1-q\right)\right]^{-1}\ \mathrm{and}\ \theta=\frac{V_{\mathrm{th}}}{\sigma_{\ex}\Gamma\sqrt{Q}},\ \tilde{\theta}=0
\end{equation}
We find that for $\alpha<\alpha_{\mathrm{b}}$ a solution only exist
for equations of the balanced case while for $\alpha>\alpha_{\mathrm{b}}$
a solution exist for the equations for the unbalanced case. Thus typical
solutions are balanced below $\alpha_{\mathrm{b}}$ and unbalanced
above it. The norm of the weight and the imbalance index depicted
in Fig. 2b and Fig. 2c are given by
\begin{equation}
\wnorm=\frac{1}{\sqrt{Q}\theta}
\end{equation}
\begin{equation}
\mathrm{IB}=\frac{\sum_{\pm}f_{\pm}\phi_{\pm}\Phi_{\pm}^{\prime}\left(B\phi_{\pm},Z,q\right)}{\sum_{\pm}\pm f_{\pm}\phi_{\pm}\Phi_{\pm}^{\prime}\left(B\phi_{\pm},Z,q\right)}
\end{equation}

\subsubsection{Solutions with significant $\protect\kout$ are balanced}

In this section we show that all unbalanced solutions have output
robustness of order $1/\sqrt{N}$ and, equivalently, Solutions with
$\kout$ of order 1 are balanced. 

\textbf{Theorem:} All Unbalanced solutions have output robustness
of the order of $1/\sqrt{N}.$ 

\textbf{Proof: }In the case of output robustness we $\tilde{K}=\tilde{K}_{\mathrm{out}}=\theta\kout$
and thus, $\frac{\partial\mathcal{F}_{h}}{\partial Q}=0$. We are
looking for unbalanced solutions ($\tilde{\theta}>0,\ \theta\sim\order{\sqrt{N}}$)
so we have the equations,
\begin{align}
\frac{\sigma_{\mathrm{exc}}B}{\bar{x}_{\mathrm{exc}}\sqrt{2CN}} & =-\frac{1}{2\left(1-q\right)}\frac{\partial\mathcal{F}_{h}}{\partial\Delta}\left(\frac{\partial\mathcal{F}_{h}}{\partial q}\right)^{-1}\label{eq:delta_eq_@}\\
\frac{\sigma_{\mathrm{exc}}B}{\bar{x}_{\mathrm{exc}}\sqrt{2CN}}= & \frac{1-q}{2\theta C}+\frac{\mathrm{1}}{2\left(1-q\right)}\frac{\partial\mathcal{F}_{h}}{\partial\theta}\left(\frac{\partial\mathcal{F}_{h}}{\partial q}\right)^{-1}\ .\label{eq:theta_t}
\end{align}
Both equations must be satisfied therefore we have (using (\ref{eq:alpha_t}),
(\ref{eq:delta_eq_@}) and (\ref{eq:theta_t}))
\begin{equation}
0=\frac{\partial\mathcal{F}_{h}}{\partial\Delta}+\frac{\partial\mathcal{F}_{h}}{\partial\theta}-\frac{1}{\alpha\theta}
\end{equation}

Performing the derivatives we get
\begin{equation}
0=\left(\kout+1\right)p_{\mathrm{out}}\int DtJ_{2}\left(-X_{+}\right)+\left(\kout-1\right)\left(1-p_{\mathrm{out}}\right)\int DtJ_{2}\left(X_{-}\right)-\frac{\sqrt{1-q}}{\alpha\theta}
\end{equation}
Now, we use eq. \ref{eq:delta_eq_@}, leading to
\begin{equation}
p_{\mathrm{out}}\int DtJ_{2}\left(-X_{+}\right)=\left(1-p_{\mathrm{out}}\right)\int DtJ_{2}\left(X_{-}\right)-M/\sqrt{N}
\end{equation}
where we defined $M$ as 
\begin{equation}
M=\frac{\sqrt{2}B\sigma_{\mathrm{exc}}\sqrt{1-q}}{\sqrt{C}\bar{x}_{\mathrm{exc}}}\frac{\partial\mathcal{F}_{h}}{\partial q}
\end{equation}
which remains of $\order 1$. Thus, we are left with
\begin{equation}
0=2\kout\left(1-p_{\mathrm{out}}\right)\int DtJ_{2}\left(X_{-}\right)-\left(\kout+1\right)\frac{M}{\sqrt{N}}-\frac{\sqrt{1-q}}{\alpha\theta}
\end{equation}
Note that $J_{2}\left(x\right)<0$ and the first term is negative
(non zero). The other two terms scale as $1/\sqrt{N}$ and therefore
the equation can be satisfied only if $\kappa=\frac{\kappa_{0}}{\sqrt{N}}$
$\blacksquare$ 

\subsubsection{Saddle point equations for critical solutions}

To find the capacity, balanced capacity and solutions with maximal
output and input robustness we consider the limit $q\rightarrow1$. 

We define,

\begin{equation}
G_{Q}=\lim_{q\rightarrow1}\frac{Q}{\left(1-q\right)}\left(\frac{\partial\mathcal{F}_{h}}{\partial q}\right)^{-1}\frac{\partial\mathcal{F}_{h}}{\partial Q}\ ,
\end{equation}
thus in this limit $Z_{\pm}$ is given by:

\begin{equation}
Z_{\pm}=\left[1-\frac{B\tilde{\theta}}{\sqrt{2C}}+\left(1-\frac{Q}{\lambda_{\pm}^{2}}\right)G_{Q}\right]^{-1}\ .
\end{equation}
In addition, 
\begin{align}
\lim_{q\rightarrow1}\Phi_{\pm}\left(x,z,q\right) & =z\gamma_{\pm}\left(x\right)\\
\gamma_{\pm}\left(x\right) & =\frac{x^{2}+1}{2}-\frac{1}{2}\int Du\left(x+u\right)^{2}\Theta\left[\pm\left(x+u\right)\right]\\
\lim_{q\rightarrow1}\Phi_{\pm}^{\prime}\left(x,z,q\right) & =z\gamma_{\pm}^{\prime}\left(x\right)\\
\gamma_{\pm}^{\prime}\left(x\right) & =x-\int Du\left(x+u\right)\Theta\left[\pm\left(x+u\right)\right]\ ,
\end{align}
and, in the $q\rightarrow1$ limit we have:
\begin{align}
\left(1-q\right)\frac{\partial\mathcal{F}_{h}}{\partial\Delta} & =M\left(\Delta,\tilde{K}\right)\\
M\left(\Delta,\tilde{K}\right) & =p_{\mathrm{out}}\int_{-\infty}^{\Delta+\tilde{K}}Dt\left(t-\Delta-\tilde{K}\right)+\left(1-p_{\mathrm{out}}\right)\int_{\Delta-\tilde{K}}^{\infty}Dt\left(t-\Delta+\tilde{K}\right)\\
\left(1-q\right)^{2}\frac{\partial\mathcal{F}_{h}}{\partial q} & =-\frac{1}{2}\alpha^{\mathrm{Unconst.}}\left(\Delta,\tilde{K}\right)\\
\alpha^{\mathrm{Unconst.}}\left(\Delta,\tilde{K}\right) & =\left[p_{\mathrm{out}}\int_{-\infty}^{\Delta+\tilde{K}}Dt\left(t-\Delta-\tilde{K}\right)^{2}+\left(1-p_{\mathrm{out}}\right)\int_{\Delta-\tilde{K}}^{\infty}Dt\left(t-\Delta+\tilde{K}\right)^{2}\right]^{-1}\ .
\end{align}
We now write the final form of the saddle point equations for critical
solutions: 
\begin{align}
C & =\sum_{\pm}f_{\pm}Z_{\pm}^{2}\gamma_{\pm}\left(B\phi_{\pm}\right)\label{eq:C_critical}\\
Q & =\frac{\sum_{\pm}f_{\pm}Z_{\pm}^{2}\gamma_{\pm}\left(B\phi_{\pm}\right)}{\sum_{\pm}\frac{f_{\pm}}{\lambda_{\pm}^{2}}Z_{\pm}^{2}\gamma_{\pm}\left(B\phi_{\pm}\right)}\\
\tilde{\theta} & =-\frac{\sum_{\pm}f_{\pm}\phi_{\pm}Z_{\pm}\gamma_{\pm}^{\prime}\left(B\phi_{\pm}\right)}{\sqrt{2\sum_{\pm}f_{\pm}Z_{\pm}^{2}\gamma_{\pm}\left(B\phi_{\pm}\right)}}\\
0 & =M\left(\Delta,\tilde{K}\right)\label{eq:Delta_final-1}\\
\frac{B}{\sqrt{2CN}} & =-\frac{\bar{x}_{\mathrm{ex}}}{\sigma_{\mathrm{ex}}}2C\alpha^{\mathrm{Unconst.}}\left(\Delta,\tilde{K}\right)\lim_{q\rightarrow1}\left(1-q\right)\frac{\partial\mathcal{F}_{h}}{\partial\theta}\ \mathrm{\boldsymbol{OR}}\ \theta=\frac{V_{\mathrm{th}}}{\sigma_{\ex}\Gamma\sqrt{Q}}\ .\label{eq:theta-final-1}
\end{align}

Finally $\alpha$ is given by:
\begin{equation}
\alpha=2C\alpha^{\mathrm{Unconst.}}\left(\Delta,\tilde{K}\right)
\end{equation}

\subsubsection{Capacity and balanced capacity}

The capacity is given for $\delta=0$ and $\kappa=0$. In this case
both $\frac{\partial\mathcal{F}_{h}}{\partial\theta}$ and $\frac{\partial\mathcal{F}_{h}}{\partial Q}$
are zero and equation \ref{eq:theta-final-1} has two possible solutions:
\begin{equation}
B=0,\ \theta>\frac{V_{\mathrm{th}}}{\sigma_{\ex}\Gamma\sqrt{Q}}\ ,
\end{equation}
for unbalanced solutions \textbf{or}
\begin{equation}
\theta=\frac{V_{\mathrm{th}}}{\sigma_{\ex}\Gamma\sqrt{Q}},\ \tilde{\theta}=0\ ,
\end{equation}
for balanced solutions. In both cases we have $Z_{\pm}=1$.

\paragraph{Unbalanced solution}

The saddle point equations become
\begin{equation}
Q=\frac{1}{\sum_{\pm}\frac{1}{\lambda_{\pm}^{2}}f_{\pm}}
\end{equation}
\begin{equation}
\tilde{\theta}=\frac{1}{\sqrt{\pi}}\sum_{\pm}\pm f_{\pm}\phi_{\pm}\label{eq:theta_tilde_capacity-1}
\end{equation}
\begin{equation}
C=\frac{1}{4}
\end{equation}
and the capacity is given by:
\begin{equation}
\alpha_{\mathrm{c}}=\frac{1}{2}\alpha^{\mathrm{Unconst.}}\left(\Delta,0\right)
\end{equation}
where $\Delta$ is given by $M\left(\Delta,0\right)=0$. 

This solution is valid only when $\theta$ is larger than its $\order 1$
lower bound which is guarantied in the large $N$ limit as long as
$\tilde{\theta}>0$. Using eq. \ref{eq:theta_tilde_capacity-1}, this
entails that
\begin{equation}
f_{\ex}>f_{\ex}^{\star}
\end{equation}
with
\begin{equation}
f_{\ex}^{\star}=\frac{\phi}{1+\phi}
\end{equation}
or conversely $\phi<\frac{f_{\mathrm{exc}}}{1-f_{\mathrm{exc}}}\ .$

\paragraph{Balanced solution}

In this solution we have $\theta=\frac{V_{\mathrm{th}}}{\sigma_{\ex}\Gamma\sqrt{Q}},\ \tilde{\theta}=0$.

$B$ is given by the solution to 
\begin{equation}
\sum_{\pm}f_{\pm}\phi_{\pm}\gamma_{\pm}^{\prime}\left(B\phi_{\pm}\right)=0\label{eq:balance_capacity-1}
\end{equation}
and we have
\begin{equation}
Q=\frac{\sum_{\pm}f_{\pm}\gamma_{\pm}\left(B\phi_{\pm}\right)}{\sum_{\pm}\frac{1}{\lambda_{\pm}^{2}}f_{\pm}\gamma_{\pm}\left(B\phi_{\pm}\right)}
\end{equation}
\begin{equation}
C=\sum_{\pm}f_{\pm}\gamma_{\pm}\left(B\phi_{\pm}\right)
\end{equation}
and 
\begin{equation}
\alpha_{\mathrm{b}}=2C\alpha^{\mathrm{Unconst.}}\left(\Delta,0\right)
\end{equation}
where $\Delta$ is given by $M\left(\Delta,0\right)=0$. 

This gives the balanced capacity line. For $f_{\ex}<f_{\ex}^{\star}$
this is the capacity line as well. Thus, for $f_{\ex}<f_{\ex}^{\star}$,
at capacity the solution is balanced. 

\subsubsection{Coexistence of balanced and unbalanced solutions below the balanced
capacity line}

To show that unbalanced solutions coexist with balanced solutions
for any $\alpha<\alpha_{\mathrm{b}}$, we calculate the capacity of
unbalanced solutions with a given norm. This can be done by solving
equations (\ref{eq:C_critical})-(\ref{eq:Delta_final-1}) while imposing
the condition $\wnorm=\frac{V_{\mathrm{thr}}W}{\sqrt{N}\sigma_{\ex}}$
through the saddle point equation of $\theta$: 
\begin{equation}
\theta=\frac{\sqrt{N}}{W\sqrt{Q}}\ .
\end{equation}
We therefor have:
\begin{equation}
\tilde{\theta}=\frac{\sigma_{\ex}\left(\theta-\Delta\right)}{\bar{x}_{\ex}\sqrt{N}}\simeq\frac{\sigma_{\ex}}{\bar{x}_{\ex}W\sqrt{Q}}=\frac{1}{\tilde{W}\sqrt{Q}}
\end{equation}
We are interested in the capacity and therefor we take $K=0$. As
a results we have:
\begin{equation}
Z_{\pm}=\left[1-\frac{B}{\sqrt{2CQ\tilde{W}^{2}}}\right]^{-1}
\end{equation}
and the saddle point equations become:
\begin{align}
\frac{1}{\tilde{W}}= & -\frac{\sum_{\pm}f_{\pm}\phi_{\pm}\gamma_{\pm}^{\prime}\left(B\phi_{\pm}\right)}{\sqrt{2\sum_{\pm}\frac{f_{\pm}}{\lambda_{\pm}^{2}}\gamma_{\pm}\left(B\phi_{\pm}\right)}}\\
C= & \left[1-\frac{B}{\tilde{W}\sqrt{2CQ}}\right]^{-2}\sum_{\pm}f_{\pm}\gamma_{\pm}\left(B\phi_{\pm}\right)\\
0= & M\left(\Delta,0\right)
\end{align}
where $Q$ is given by
\begin{equation}
Q=\frac{\sum_{\pm}f_{\pm}\gamma_{\pm}\left(B\phi_{\pm}\right)}{\sum_{\pm}\frac{f_{\pm}}{\lambda_{\pm}^{2}}\gamma_{\pm}\left(B\phi_{\pm}\right)}
\end{equation}
Given the value of $B$ and $Q$ the equation for $C$ can be solved
for $\sqrt{C}$ and we get: 

\begin{equation}
\sqrt{C}=\frac{B}{\sqrt{2Q}\tilde{W}}+\sqrt{\sum_{\pm}f_{\pm}\gamma_{\pm}\left(B\phi_{\pm}\right)-\left(\frac{B}{\sqrt{2Q}\tilde{W}}\right)^{2}}
\end{equation}
This solution is valid as long as $\sqrt{C}\ge0$. The conditional
capacity $\alpha_{\mathrm{c}}\left(\tilde{W}\right)$ is then given
by
\begin{equation}
\alpha_{\mathrm{c}}\left(\tilde{W}\right)=2C\alpha^{\mathrm{Unconst.}}\left(\Delta,0\right)\ .
\end{equation}

It is easy to see that for $\tilde{W}\rightarrow\infty$, the saddle
point equations converge to the equations of the balanced capacity
and thus $\alpha_{\mathrm{c}}\left(\tilde{W}\right)$ approaches $\alpha_{\mathrm{b}}$.
In addition we find that for $f_{\mathrm{exc}}<f_{\mathrm{exc}}^{\star}$,
$\alpha_{\mathrm{c}}\left(\tilde{W}\right)$ is a monotonically increasing
function of $\tilde{W}$. Another way to interpret this result is
to `invert the function' and ask what is the minimal value of $\tilde{W}$
that permits solutions given $\alpha$. Our result implies that strictly
below $\alpha_{\mathrm{b}}$ the minimal value of $\tilde{W}$ that
permits solutions is of $\order 1$ (i.e. $\wnorm$ of $\order{1/\sqrt{N}}$)
and unbalanced solutions exist. The minimal $\tilde{W}$ diverges
as $\alpha$ approaches $\alpha_{\mathrm{b}}$ and hence the solution
at $\alpha_{\mathrm{b}}$ is balanced ($\wnorm$ of $\order 1$).

\subsubsection{\label{subsec:max_kappa_in}Saddle point equations for the maximal
$\protect\kin$ solution}

In this case we have $\tilde{K}=\tilde{K}_{\text{in}}=\frac{\delta}{\sqrt{Q}}$
and therefore $\frac{\partial\mathcal{F}_{h}}{\partial\theta}=0$.
For unbalanced solutions we have
\begin{equation}
B=0,\ \theta>\frac{V_{\mathrm{th}}}{\sigma_{\ex}\Gamma\sqrt{Q}},\ \tilde{\theta}>0
\end{equation}
and for balanced solutions we have
\begin{equation}
\theta=\frac{V_{\mathrm{th}}}{\sigma_{\ex}\Gamma\sqrt{Q}},\ \tilde{\theta}=0\ .
\end{equation}

In both cases $Z_{\pm}$ is given by
\begin{equation}
Z_{\pm}=\left[1+\left(1-\frac{Q}{\lambda_{\pm}^{2}}\right)G_{Q}\right]^{-1}
\end{equation}
with
\begin{align}
G_{Q}= & \kin/\sqrt{Q}\alpha^{\mathrm{Unconst.}}\left(\Delta,\kin/\sqrt{Q}\right)\left[p_{\mathrm{out}}\int_{-\infty}^{\Delta+\frac{\kin}{\sqrt{Q}}}Dt\left(t-\Delta-\frac{\kin}{\sqrt{Q}}\right)\right.\label{eq:G_q_margin-1}\\
 & \left.-\left(1-p_{\mathrm{out}}\right)\int_{\Delta-\frac{\kin}{\sqrt{Q}}}^{\infty}Dt\left(t-\Delta+\frac{\kin}{\sqrt{Q}}\right)\right]\nonumber 
\end{align}

\paragraph{Unbalanced solution}

In this case we have equations for $\Delta$ and $Q$: 
\begin{equation}
M\left(\Delta,\kin/\sqrt{Q}\right)=0\label{eq:Delta_margin-1-1}
\end{equation}
\begin{equation}
Q=\frac{\sum_{\pm}f_{\pm}Z_{\pm}^{2}}{\sum_{\pm}\frac{1}{\lambda_{\pm}^{2}}f_{\pm}Z_{\pm}^{2}}\ .\label{eq:Q_margin-1}
\end{equation}

We then have:
\begin{equation}
\tilde{\theta}=\frac{\frac{1}{\sqrt{2\pi}}\sum_{\pm}\pm f_{\pm}\phi_{\pm}Z_{\pm}}{\sqrt{\frac{1}{2}\sum_{\pm}f_{\pm}Z_{\pm}^{2}}}
\end{equation}
\begin{equation}
C=\frac{1}{4}\sum_{\pm}f_{\pm}Z_{\pm}^{2}
\end{equation}
and 
\begin{equation}
\alpha=2C\alpha^{\mathrm{Unconst.}}\left(\Delta,\kin/\sqrt{Q}\right)\label{eq:alpha_margin-1}
\end{equation}

\paragraph{\label{subsec:Balanced-solution}Balanced solution}

In this case we have equations for $\Delta$, $B$ and $Q$:
\begin{equation}
M\left(\Delta,\kin/\sqrt{Q}\right)=0
\end{equation}
\begin{equation}
Q=\frac{\sum_{\pm}f_{\pm}\gamma_{\pm}\left(B\phi_{\pm}\right)Z_{\pm}^{2}}{\sum_{\pm}\frac{1}{\lambda_{\pm}^{2}}f_{\pm}\gamma_{\pm}\left(B\phi_{\pm}\right)Z_{\pm}^{2}}
\end{equation}
\begin{eqnarray}
0 & = & \sum_{\pm}f_{\pm}\phi_{\pm}\gamma_{\pm}^{\prime}\left(B\phi_{\pm}\right)Z_{\pm}
\end{eqnarray}
the equations for $G_{Q}$ (\ref{eq:G_q_margin-1}) and $\alpha$
(\ref{eq:alpha_margin-1}) remain the same, however $C$ is given
by:
\begin{equation}
C=\sum_{\pm}f_{\pm}\gamma_{\pm}\left(B\phi_{\pm}\right)Z_{\pm}^{2}
\end{equation}

\paragraph{Transition between balanced and unbalanced solutions}

Transition points between balanced and unbalanced solutions depend
on the value of $\phi,$ $\lambda$ and $f_{\mathrm{exc}}$. Transition
points are points in which both $B=0$ and $\tilde{\theta}=0$. Thus
we have
\begin{equation}
\phi^{\star}=\frac{f_{\mathrm{exc}}Z_{+}}{\left(1-f_{\mathrm{exc}}\right)Z_{-}}\label{eq:phi_star-1}
\end{equation}
where $Q$ and $\Delta$ are given by (\ref{eq:Q_margin-1}) and (\ref{eq:Delta_margin-1-1}).
Thus $\phi^{\star}$ is a function of $\kin$ and $\lambda$. Solutions
are balanced for $\phi>\phi^{\star}$ and unbalanced for $\phi<\phi^{\star}$
{[}Fig. \ref{fig:input_stats}b{]}. 

\subsubsection{Saddle point equations for the maximal $\protect\kout$ solution}

\paragraph{Unbalanced solution}

This solution is valid for $\alpha>\alpha_{\mathrm{b}}$, $f_{\ex}>f_{\ex}^{\star}$.
We look for a solution with $\tilde{\theta}>0$ thus $\theta$ must
scale as $\sqrt{N}$.

In this case $\tilde{K}=\theta\kappa$ and so $\frac{\partial\mathcal{F}_{h}}{\partial Q}=0$
and $Z_{\pm}=\left[1-\frac{B\tilde{\theta}}{\sqrt{2C}}\right]^{-1}$. 

We then have: 
\begin{equation}
Q=\frac{\sum_{\pm}f_{\pm}\gamma_{\pm}\left(B\phi_{\pm}\right)}{\sum_{\pm}\frac{1}{\lambda_{\pm}^{2}}f_{\pm}\gamma_{\pm}\left(B\phi_{\pm}\right)}
\end{equation}
And we are left with equations to solve for $\tilde{\theta},$ $B$
and $\Delta$. 
\begin{equation}
\tilde{\theta}=-\frac{\sum_{\pm}f_{\pm}\phi_{\pm}\gamma_{\pm}^{\prime}\left(B\phi_{\pm}\right)}{\sqrt{2\sum_{\pm}f_{\pm}\gamma_{\pm}\left(B\phi_{\pm}\right)}}
\end{equation}

\begin{equation}
M\left(\Delta,\tilde{K}\right)=0
\end{equation}
and
\begin{equation}
\frac{B}{\sqrt{2CN}}=-\frac{\bar{x}_{\mathrm{ex}}}{\sigma_{\mathrm{ex}}}\kappa\alpha^{\mathrm{Unconst.}}\left(\Delta,\tilde{K}\right)\left[f\int_{-\infty}^{\Delta+\tilde{K}}Dt\left(t-\Delta-\tilde{K}\right)-\left(1-f\right)\int_{\Delta-\tilde{K}}^{\infty}Dt\,\left(t-\Delta+\tilde{K}\right)\right]
\end{equation}
There is only a solution if 
\begin{equation}
\kappa=\frac{\sigma_{\ex}}{\bar{x}_{\ex}\sqrt{N}}\kappa_{0},\ \theta\simeq\sqrt{N}\frac{\bar{x}_{\mathrm{ex}}}{\sigma_{\mathrm{ex}}}\tilde{\theta}
\end{equation}
so we have $\tilde{K}=\tilde{\theta}\kappa_{0}$ and 
\begin{equation}
\frac{B}{\sqrt{2C}}=-\kappa_{0}\alpha^{\mathrm{Unconst.}}\left(\Delta,\tilde{\theta}\kappa_{0}\right)\left[f\int_{-\infty}^{\Delta+\tilde{\theta}\kappa_{0}}Dt\left(t-\Delta-\tilde{\theta}\kappa_{0}\right)-\left(1-f\right)\int_{\Delta-\tilde{\theta}\kappa_{0}}^{\infty}Dt\,\left(t-\Delta+\tilde{\theta}\kappa_{0}\right)\right]\ .
\end{equation}
Finally we have $C=Z^{2}\sum_{\pm}f_{\pm}\gamma_{\pm}\left(B\phi_{\pm}\right),$
from which we can isolate $C$ to have
\begin{equation}
C=\left[\sum_{\pm}f_{\pm}\gamma_{\pm}\left(B\phi_{\pm}\right)\right]\left[1-\frac{B\sum_{\pm}f_{\pm}\phi_{\pm}\gamma_{\pm}^{\prime}\left(B\phi_{\pm}\right)}{2\sum_{\pm}f_{\pm}\gamma_{\pm}\left(B\phi_{\pm}\right)}\right]^{2}\ .
\end{equation}
 $\alpha$ is given as before $\alpha=2C\alpha^{\mathrm{Unconst.}}\left(\Delta,\tilde{\theta}\kappa_{0}\right)$. 

The equations given in this section are equivalent to the ones derived
in \cite{chapeton_efficient_2012}.

\paragraph{Balanced solution }

We look for balanced solutions with $\theta=\frac{V_{\mathrm{th}}}{\sigma_{\ex}\Gamma\sqrt{Q}},\ \tilde{\theta}=0$.
The saddle point equations in this case are given by the same equations
as the balanced solution described in subsection \ref{subsec:Balanced-solution}
with $\kout V_{\mathrm{th}}/\sigma_{\ex}\Gamma$ replacing $\kin$. 

\subsubsection{\label{subsec:Distribution-of-synaptic}Distribution of synaptic
weights}

We derive the mean distribution of synaptic weights for critical solutions
($q\rightarrow1$)
\begin{equation}
P_{\pm}\left(w\right)=H\left(\mp B\phi_{\pm}\right)\delta\left(w\right)+\sqrt{\frac{\theta^{2}N\lambda_{\pm}^{2}}{2\pi\sigma_{w\pm}^{2}}}\exp\left[-\frac{\left(\theta\sqrt{N}\lambda_{\pm}w+B\phi_{\pm}\sigma_{w\pm}\right)^{2}}{2\sigma_{w\pm}^{2}}\right]\ ,
\end{equation}
with $\sigma_{w\pm}=\frac{Z_{\pm}}{\sqrt{2C}}$, where $P_{+}$ and
$P_{-}$ denote the probability densities for excitatory and inhibitory
synaptic weights respectively, $\delta\left(x\right)$ is the Dirac
delta function and weights are given in units of $V_{\mathrm{th}}/\sigma_{\mathrm{exc}}$.
The fraction of silent synapses is given by $H\left(-B\right)$ for
excitatory synapses and by $H\left(B\phi\right)$ for inhibitory synapses.

\end{document}